\def\mode{1}
\if 0\mode
  \documentclass[12pt,onecolumn,draftclsnofoot,letterpaper]{IEEEtran}%draftcls,onecolumn,12pt
  \let\chapter\section
\else
	\documentclass[journal]{IEEEtran}
\fi
\usepackage{graphicx}
\usepackage{amsmath}
\usepackage{amssymb}
\usepackage{multirow}
\usepackage{tabularx}
\usepackage{listliketab}
\usepackage{cite}
\usepackage{url}
\usepackage{color}
\usepackage{amsthm}

\usepackage{algorithm}
\usepackage{algpseudocode}
\usepackage{algorithmicx}
\usepackage{pifont}
\usepackage{epstopdf} 
\usepackage{bm}
\newtheorem*{theorem*}{Theorem}
\usepackage{enumitem}
\newcommand{\mat}{\mathbf}
\renewcommand{\t}[1]{\text{#1}}
\renewcommand{\v}[1]{\bar{\mathbf{#1}}}
\newcommand{\C}[1]{\mathcal{#1}}
\renewcommand{\vec}[1]{\bar{\pmb{#1}}}
\ifCLASSOPTIONcompsoc
\usepackage[caption=false,font=normalsize,labelfont=sf,textfont=sf]{subfig}
\else
\usepackage[subrefformat=parens,labelformat=parens,caption=false,font=footnotesize]{subfig}
\fi

\usepackage{array}
\newcolumntype{L}[1]{>{\raggedright\let\newline\\\arraybackslash\hspace{0pt}}m{#1}}
\newcolumntype{C}[1]{>{\centering\let\newline\\\arraybackslash\hspace{0pt}}m{#1}}
\newcolumntype{R}[1]{>{\raggedleft\let\newline\\\arraybackslash\hspace{0pt}}m{#1}}

\graphicspath{{./figures/}}

\ifCLASSINFOpdf

\else

\fi

\usepackage{cleveref}
\begin{document}

\title{Optimal User Scheduling in Energy Harvesting\\ Wireless Networks}

\author{Kalpant~Pathak$^*$,~\IEEEmembership{Student Member,~IEEE,}
        Sanket~S.~Kalamkar$^{\dagger}$,~\IEEEmembership{Member,~IEEE,}
        and~Adrish~Banerjee$^*$,~\IEEEmembership{Senior~Member,~IEEE}% <-this % stops a space
\thanks{$^*$Authors are with the Department of Electrical Engineering, Indian Institute of Technology Kanpur, UP, 208016 India, e-mail: \{kalpant, adrish\}@iitk.ac.in.\newline \indent$^{\dagger}$Author is with the Department of Electrical Engineering, University of Notre Dame, IN, 46556, USA, e-mail: skalamka@nd.edu.\newline
\indent This work was carried out while Sanket.~S.~Kalamkar was at Indian Institute of Technology Kanpur, India.}% <-this % stops a space
}

\maketitle

\begin{abstract}
We consider a wireless network where multiple energy harvesting transmitters communicate with the common receiver in a time sharing manner. In each slot, a transmitter can either harvest energy or send its data to the receiver. Given a time deadline, the goal is to maximize the sum-rate of transmitters under random energy arrivals with both perfect and imperfect channel state information (CSI) at the receiver. The original sum-rate maximization (SRM) problem is a non-convex mixed integer non-linear program (MINLP). To obtain the optimal scheduling policy, we first reduce the original optimization problem to a convex MINLP and solve it using the generalized Benders decomposition algorithm. We observe that the SRM problem results in an unfair rate allocation among transmitters, \textit{i.e.}, the transmitter closer to the receiver achieves a higher rate than that by the transmitter farther from the receiver. Hence, to induce fairness among transmitters, we consider the minimum-rate maximization (MRM) problem. For the bounded channel estimation error, we obtain a robust scheduling policy by solving the worst-case SRM and MRM problems. Finally, we compare the proposed policies with myopic policies studied in the literature and show that the former outperform the latter in terms of achievable rates.
\end{abstract}
%We compare both the scheduling policies and observe that the \textit{sum-rate maximization} policy outperforms the \textit{minimum-rate maximization} policy, however latter maintains the fairness among the users.

\IEEEpeerreviewmaketitle
\section{Introduction}
\label{sec:introduction}
\IEEEPARstart{I}{n} recent years, energy harvesting (EH) in wireless networks has emerged as a promising technology to achieve a sustained and low-cost operation of communication devices \cite{EH_survey_Ku,EH_survey_1,EH_survey_2,EH_survey_Sudevalayam,swipt_magzine_krikidis}. These devices obtain energy from environmental sources such as solar power, vibration, radio-frequency (RF) signals, etc. The sporadic nature of energy arrival demands designing optimal resource allocation policies to optimize the energy utilization and the system performance simultaneously. In multiple access networks such as wireless sensor networks, multiple transmitters wish to communicate with a common receiver. However, due to varying channel conditions, the network throughput is significantly affected by the order in which the transmitters access the channel. In addition, if the transmitters have energy harvesting capability, the network throughput depends on the energy availability at transmitters. Hence, the design of optimal user scheduling and power control policies is important to maximize the network throughput in EH multiple access networks.
\begin{table*}[!h]
	\centering
	\caption{Background works}
	\begin{tabular}{|C{2.6cm}|C{1cm}|C{4cm}|C{5.5cm}|C{1.3cm}|}
		\hline
		\textbf{Literature} & \textbf{EH} & \textbf{System Model} & \textbf{Policy} & \textbf{CSI}\\
		\hline
		J.~Yang \emph{et al.} \cite{TCTM_ulukus} & Generic & Point-to-point link, no fading, random data and energy arrivals & Transmission completion time minimization (TCTM) & Not required\\
		\hline
		K.~Tutuncuoglu \emph{et al.} \cite{STTM_yener} & Generic & Point-to-point link, no fading, random energy arrivals, Finite battery & Short-term throughput maximization (STTM) & Not required\\
		\hline
		O.~Ozel \emph{et al.} \cite{STTM_TCTM_fading} & Generic & Point-to-point link, random data and energy arrivals & STTM and TCTM, optimal offline and online policies & Perfect \\
		\hline
		H.~Ju \emph{et al.} \cite{TDMA_zhang} & RF & Half-duplex HAP, TDMA & Optimal time allocation, myopic policy & Perfect\\
		\hline
		Z.~Hadzi-Velkov \emph{et al.} \cite{TDMA_velkov} & RF & Half-duplex HAP, TDMA & Joint optimization of HAP power and time allocation, myopic policy & Perfect\\
		\hline
		X.~Kang \emph{et al.} \cite{TDMA_full_duplex_kang} & RF & Full-duplex HAP, TDMA & Optimal time allocation, myopic policy & Perfect\\
		\hline
		D.~Xu \emph{et al.} \cite{TDMA_underlay_CRN} & RF & CRN with multiple EH-SUs, TDMA & Joint optimization of BS power and time allocation, myopic policy & Perfect\\
		\hline
		H.~Ju \emph{et al.}\cite{full_duplex_zhang} & RF & Full-duplex HAP, TDMA& Joint optimization of HAP power and time allocation, myopic policy & Perfect \\
		\hline
		Q.~Wu \emph{et al.}\cite{TDMA_vs_NOMA} & RF & Harvest from PB, transmit to AP, circuit power consumption, TDMA, NOMA & Optimal time allocation in TDMA and NOMA, myopic policy & Perfect \\
		\hline
		I.~Ahmed \emph{et al.} \cite{imperfect_CSI_relay_ahmed} & Generic & Two-way half-duplex DF relay, multiple access and time division broadcast relaying & Robust joint energy and transmission time allocation, optimal offline & Imperfect \\
		\hline
		I.~Ahmed \emph{et al.} \cite{imperfect_CSI_relay_half_duplex_ahmed} & Hybrid & Half-duplex DF relay, energy state uncertainty & Robust power allocation, optimal offline, optimal online, and suboptimal online & Imperfect \\
		\hline
		S.~Gong \emph{et al.} \cite{imperfect_CSI_CRN} & Generic & Underlay CRN, energy state uncertainty & Robust power control of SUs under worst case interference constraint of PUs & Imperfect \\
		\hline
		E.~Boshkovska \emph{et al.} \cite{robust_RA_MIMO} & RF & EH MIMO users communicate with a common receiver using TDMA & Joint time allocation and power control, myopic policy & Imperfect\\
		\hline
		J.~Xiao \emph{et al.} \cite{robust_transceiver_MIMO_qin} & RF & Two-user EH MIMO interference channel with SWIPT & Robust transceiver design & Imperfect\\
		\hline
		T. Peng \emph{et al.} \cite{robust_transceiver_swipt_wang} & RF & EH-MISO interference channel with SWIPT & Robust transmit beamforming and power splitting & Imperfect\\
		\hline
	\end{tabular}
	\label{table:background_works}
\end{table*}
\subsection{Background Work}
\label{sec:background}
Transmission and scheduling policies in EH wireless networks have been studied extensively in the literature. Optimal transmission policies for a point-to-point link with infinite energy buffer were proposed for random energy arrivals in \cite{TCTM_ulukus,STTM_yener}, for random data and energy arrivals in \cite{TCTM_ulukus}, and for fading channels in \cite{STTM_TCTM_fading}. The works in \cite{TDMA_zhang,TDMA_velkov,TDMA_full_duplex_kang,TDMA_underlay_CRN,full_duplex_zhang} considered the design of optimal scheduling and transmission policies for multi-user EH networks assuming the perfect channel state information (CSI). In an RF energy harvesting setting, the authors in \cite{TDMA_zhang} considered a time-division multiple access (TDMA) network with a half-duplex hybrid access point (HAP). The communication period was divided into an EH phase and a transmission phase. In the EH phase, all transmitters harvested RF energy from the constant-power signal transmitted by the HAP and then transmitted their data to the HAP in the transmission phase following the TDMA protocol in a fixed order of their distances from the HAP. Each transmitter followed a myopic transmission policy, where it consumed all the harvested energy for the transmission. Under these settings, authors obtained an optimal time allocation policy that maximized the sum-rate and the minimum rate among the transmitters. In \cite{TDMA_velkov}, the authors proposed a myopic scheduling policy that outperformed the one presented in \cite{TDMA_zhang} by jointly optimizing the HAP's transmit power and the transmission time allocation. The authors considered the sum-rate maximization problem only, where the transmitters sent their data in a fixed order using TDMA. In \cite{TDMA_full_duplex_kang}, the authors extended the system model of \cite{TDMA_zhang} to a full-duplex HAP where it could transmit RF energy and receive information simultaneously. During the transmission phase, when a transmitter was transmitting, the transmitters that were scheduled to transmit later harvested the energy transmitted by the HAP. In this way, the transmitters farther from the HAP had higher energies than those in \cite{TDMA_zhang}, which improved the sum-rate and the fairness among transmitters. However the authors considered only the sum-rate maximization problem. In \cite{TDMA_underlay_CRN}, the authors considered an underlay cognitive radio network (CRN) following the fixed-order TDMA protocol. Similar to the harvest-then-transmit model of \cite{TDMA_velkov}, authors jointly optimized the transmit power of the base station (BS) and the time allocation among secondary users (SUs) transmitting according to a myopic policy. In \cite{full_duplex_zhang}, the authors considered a similar model to \cite{TDMA_full_duplex_kang}. However the authors jointly optimized the transmit power of the HAP and time allocation among transmitters. In the direction of previous works on fixed-order TDMA, the authors in \cite{TDMA_vs_NOMA} considered a system model inspired by \cite{TDMA_zhang}, where multiple users harvested RF energy from the power beacon (PB) and communicated with an access point (AP) using TDMA and non-orthogonal multiple access (NOMA) protocols. It was shown that, for energy-limited networks, the spectral efficiency (SE) of both protocols is the same; in fact, when the power consumption by the circuitry is not neglected, the TDMA outperformed the NOMA with a significant gap in terms of energy consumption and SE.

The aforementioned works on scheduling policies for EH networks considered only a fixed transmission order with myopic policies and perfect CSI. However, the presence of the noise makes it difficult to estimate channel coefficients perfectly. Thus the study of EH wireless networks under channel estimation errors is important. Robust offline resource allocation policies for the decode-and-forward (DF) relay network were proposed in \cite{imperfect_CSI_relay_ahmed} for the multiple access broadcast channel and in \cite{imperfect_CSI_relay_half_duplex_ahmed} for a half-duplex relay. In \cite{imperfect_CSI_CRN}, the authors considered the distribution uncertainty model for imperfect CSI, where the distribution of channel coefficients was unknown but had a finite divergence from an empirical distribution. Under this setting, the authors obtained a robust power allocation policy for SUs in an underlay CRN. In \cite{robust_RA_MIMO}, the authors considered a non-linear EH model and obtained a robust time and power allocation policy in a wireless powered TDMA-MIMO communication network with a single-slot setting. In \cite{robust_transceiver_MIMO_qin}, the authors considered a two-user EH multiple input multiple output (MIMO) interference channel with simultaneous wireless information and power transfer (SWIPT). The authors proposed a robust transceiver design considering the bounded channel uncertainty model. In \cite{robust_transceiver_swipt_wang}, the authors considered a multiple-input single-output (MISO) interference channel where multiple multi-antenna transmitters communicated with their respective single-antenna receivers. Using the bounded uncertainty model, the authors obtained a robust transmit beamforming and power splitting that minimized the total transmit power. Table \ref{table:background_works} summarizes the background work.
\subsection{Motivation} 
\label{sec:contribution}
While communicating over a fading channel, a myopic policy may not be optimal for an EH transmitter due to varying channel conditions; rather saving energy for future slots might improve the throughput. Also apart from time sharing, the scheduling order of transmitters, which plays an important role in determining the performance of multi-user networks \cite{TDMA_full_duplex_kang}, has not been studied in the literature so far. Therefore we propose a scheduling policy where the finite number of available slots are optimally distributed among transmitters. In addition, the proposed policy considers the power optimization in the sense that the harvested energy might not be used completely in a single transmission. The amount of energy used depends on channel conditions and the energy harvested by other transmitters. Different from previous works in \cite{TDMA_zhang,TDMA_velkov,TDMA_full_duplex_kang,TDMA_underlay_CRN,full_duplex_zhang}, in the proposed scheduling policy, a slot is not shared among the harvesting and transmission phases and its length remains fixed; rather, for each transmitter, a slot is dedicated to either harvesting energy from the environment or transmitting its data depending on the energy availability and channel conditions. Also, in the scheduling policies of \cite{TDMA_zhang,TDMA_velkov}, and \cite{TDMA_underlay_CRN}, a transmitter gets a single chance to harvest energy, and once the energy harvesting phase is over, it remains idle until its turn for transmission. On the other hand, in our policy, a transmitter may transmit multiple times based on its energy availability and channel conditions. Also a transmitter harvests energy when an another transmitter is transmitting. This strategy improves the system performance by allowing a transmitter to accumulate more energy. The design of the optimal order scheduling and power control policy requires the CSI to be available at the receiver. However the channel estimation techniques are prone to error, and therefore we also consider the effect of imperfect CSI on the proposed policies, which is not studied in \cite{TDMA_zhang,TDMA_velkov,TDMA_full_duplex_kang,TDMA_underlay_CRN,full_duplex_zhang}.

\subsection{Contributions}
We propose a joint slot allocation and power control policy in a multi-user time sharing network where a number of energy harvesting transmitters wish to communicate with the common receiver (Rx). We assume that the transmitters operate in an energy half-duplex mode as in \cite{save_then_transmit_zhang} and follow a \textit{harvest-or-transmit} protocol, \textit{i.e.}, in a slot, a transmitter can either harvest energy or transmit its data to the receiver. In addition, we consider the case where the CSIs of links between the transmitters and the receiver are imperfect with bounded uncertainties. The receiver first estimates channel coefficients using the pilot symbols sent by the transmitters. Then it obtains the optimal scheduling policy using the estimated channel gains and the energy information of  transmitters and broadcasts the evaluated policy in the downlink using an ideal backhaul. This network is a generic one, and to the best of our knowledge, it has not been considered in the literature. We aim to maximize the sum-rate of the network and the minimum rate among the transmitters by a given time deadline. The major highlights of this paper are as follows:
\begin{enumerate}
\item We first consider the problem of sum-rate maximization (SRM) by a given time deadline assuming the perfect CSI at the receiver. We formulate the problem as a mixed integer non-linear program (MINLP), which is a non-convex problem due to coupled variables. We then reduce this non-convex MINLP to a convex MINLP \cite{convex_MINLP} by decoupling the variables and obtain the optimal solution using the generalized Benders decomposition (GBD) algorithm \cite{GBD}.
\item The SRM policy results in an unfair rate allocation among transmitters. We address this issue by maximizing the minimum rate among the transmitters when the perfect CSI at the receiver is available.
\item We then consider the case of imperfect CSI at the receiver. We assume a bounded channel estimation error model and obtain a robust scheduling policy that maximizes the worst-case sum-rate and worst-case minimum rate in SRM and MRM problems, respectively. We compare the proposed joint slot allocation and power control policies and study the effects of various system parameters such as the number of slots and users, path loss exponent, and the channel estimation error.
\item  Finally, we compare the proposed SRM and MRM policies with the myopic policies proposed in \cite{TDMA_zhang,TDMA_full_duplex_kang} and show that the proposed policies outperform the myopic policies in terms of achievable rates.
\end{enumerate}

%However, the works in  \cite{TDMA_zhang} and \cite{full_duplex_sun} consider myopic policies where only the time allocation among the transmitter is optimized. 
\subsection{Paper Organization and Notation}
\label{sec:paper_organization}
The rest of the paper is organized as follows. Section~\ref{sec:sys_mod} presents the system model. Section~\ref{sec:sum_rate} discusses the SRM policy with perfect CSI. Section \ref{sec:min_rate} presents the MRM policy with perfect CSI. Sections \ref{sec:sum_rate_imp} and \ref{sec:min_rate_imp} present the SRM and MRM policies with imperfect CSI, respectively. Section \ref{sec:myopic} presents myopic policies proposed in \cite{TDMA_zhang,TDMA_full_duplex_kang}. The results are discussed in Section \ref{sec:results}. Section \ref{sec:future} discusses the future directions and Section \ref{sec:conclusions} provides the conclusions.

\textit{Notation:}  A bold-faced symbol (\textit{e.g.}, $\mat{A}$ or $\pmb{\theta}$) represents a matrix and $[\mathbf{A}]_{i,j}$ represents the entry in $i$th row and $j$th column of matrix $\mathbf{A}$. A bold-faced symbol with a ``bar'' (\textit{e.g.}, $\v{x}$ or $\vec{\delta}$) represents a vector, $\v{x}\preceq\vec{0}$ implies that every element $x_i$ of vector $\v{x}$ is less than or equal to 0, $[u]^+$ represents $\max\{u, 0\}$, $\mathbb{E}[\cdot]$ denotes the expectation operator, $\|\v{x}\|_p$ represents the $l_p$-norm, and $u\propto c$ means that $u$ is proportional to $c$. $\mathbb{R}^m$ represents a set of $m$-dimensional vectors whose elements are real numbers, while $\mathbb{R}_+^{n_1\times n_2}$ represents a set of $n_1\times n_2$ matrices such that every element of the matrix is a positive real number. Other notations used in the paper are given in Table \ref{table:notation}.
\begin{table}[t]
	\centering
		\caption{Notation}
	\begin{tabular}{|C{1cm}|L{5.5cm}|}
		\hline
		$K$ & Number of transmitters\\
		\hline
		$M$ & Number of slots\\
		\hline
		$\tau$ & Slot length\\
		\hline
		$E_0^k$ & Initial energy in the battery of $k$th Tx\\
		\hline
		$E_{H}^{k,i}$ & Energy harvested by $k$th Tx in $i$th slot\\
		\hline
		$P_{k,i}$ & Transmit power of $k$th Tx in $i$th slot\\
		\hline
		$g_{k,i}$ & Channel coefficient between $k$th Tx and receiver in $i$th slot\\
		\hline
		$h_{k,i}$ & Channel power gain between $k$th Tx and receiver in $i$th slot\\
		\hline
		$D_{k}$ & Distance between $k$th Tx and receiver\\
		\hline
		$\alpha$ & Path loss exponent\\
		\hline
		$\bar{R}$ & Target rate for the MRM policy\\
		\hline
	\end{tabular}
	\label{table:notation}
\end{table}

%\begin{table*}[!ht]
%\centering
%\begin{minipage}[h]{0.48\linewidth}
%	\captionof{table}{Notations}
%\begin{tabular}{|C{1cm}|L{5.5cm}|}
%\hline
%$K$ & Number of transmitters\\
%\hline
%$M$ & Number of slots\\
%\hline
%$\tau$ & Slot length\\
%\hline
%$E_0^k$ & Initial energy in the battery of $k$th Tx\\
%\hline
%$E_{H}^{k,i}$ & Energy harvested by $k$th Tx in $i$th slot\\
%\hline
%$P_{k,i}$ & Transmit power of $k$th Tx in $i$th slot\\
%\hline
%$g_{k,i}$ & Channel coefficient between $k$th Tx and receiver in $i$th slot\\
%\hline
%$h_{k,i}$ & Channel power gain between $k$th Tx and receiver in $i$th slot\\
%\hline
%$D_{k}$ & Distance between $k$th Tx and receiver\\
%\hline
%$\alpha$ & Path loss exponent\\
%\hline
%$\bar{R}$ & Target rate for the MRM policy\\
%\hline
%\end{tabular}
%\label{table:notation}
%\end{minipage}
%\begin{minipage}{0.48\linewidth}
%\centering
%\includegraphics[width=0.95\linewidth]{system_model.eps}
%\captionof{figure}{An energy harvesting network with a \textit{harvest-or-transmit} protocol.}
%\label{fig:system_model}
%\end{minipage}
%\end{table*}
%%%%%%%%%%%%%%%%%%%%%%%%%%%%%%%%%%%%%%%%%%%%%%%%%%%%%%%%%%%%%%%%%
\section{System Model}
\label{sec:sys_mod}
\begin{figure}[!ht]
\centering
\includegraphics[width=0.9\linewidth]{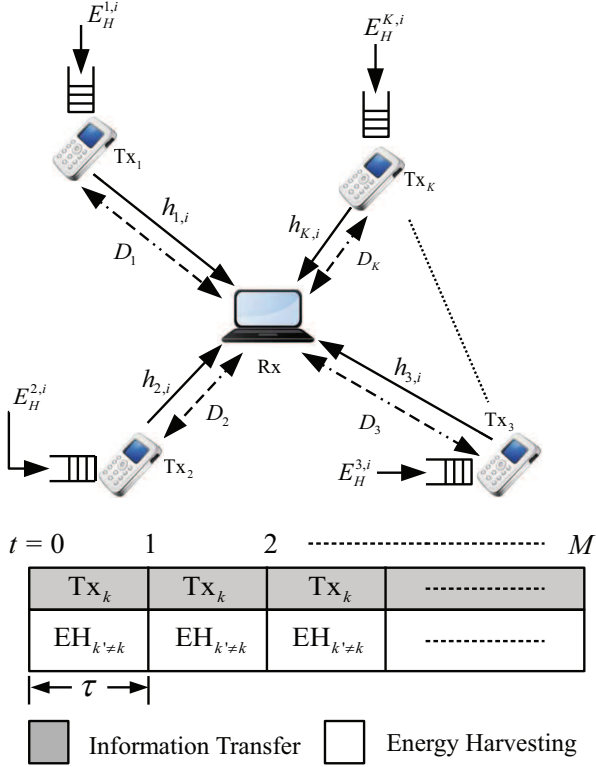}
\caption{An energy harvesting network with a \textit{harvest-or-transmit} protocol.}
\label{fig:system_model}
\end{figure}
%$E_{C}^{k,i}$ & Energy consumed by $k$th Tx in $i$th slot\\
%\hline
Fig. \ref{fig:system_model} shows an energy harvesting network consisting of $K$ energy harvesting single-antenna transmitters that communicate with a common single-antenna receiver located at a distance of $D_k$ meters from the transmitter $\text{Tx}_k$, $k=1,\ldots,K$. The transmitters send their data to the receiver over $M$ slots each of length $\tau$ seconds.

The channel coefficient $g_{k,i}$ between the $k$th transmitter and the receiver in the $i$th slot is an i.i.d. complex Gaussian random variable with zero mean and unit variance. Hence the channel power gains $h_{k,i}=|g_{k,i}|^2$ are i.i.d. exponential random variables. We assume quasi-static flat-fading where the channel coefficients $g_{k,i}$ remain constant for each transmission slot and may vary independently from one slot to another. These channel coefficients can be obtained before the transmission using a channel estimation/prediction technique \cite{channel_estimation_survey_2,fading_prediction,pilot_based_prediction_1,channel_estimation_survey}.

The proposed model employs a \textit{harvest-or-transmit} protocol where the energy arrivals are random in nature. In slot $i$, $i=1,\ldots,M$, the transmitter $\text{Tx}_{k\in\{1,\ldots,K\}}$ transmits with power $P_{k,i}$ while the transmitters Tx$_{j\in\{1,\ldots,K\}\backslash \{k\}}$ harvest energy $E_H^{j,i}, \forall j\in\{1,\ldots,K\}\backslash k$ and store it in their sufficiently large-capacity batteries. Each transmitter $\text{Tx}_{k}$ has some initial energy $E_0^k$ in its battery so that the transmission can start from the first slot itself. 

During the transmission in $i$th slot, Tx$_k$ transmits a signal $s_{k,i}$ with power $P_{k,i}=\mathbb{E}\left[|s_{k,i}|^2\right]$. The received signal at the receiver is
\begin{align*}
y_{k,i}=g_{k,i} s_{k,i}+n_{k,i},
\end{align*}
where $n_{k,i}\sim\mathcal{CN}(0,\sigma_n^2)$ with $\mathcal{CN}(\mu,\sigma^2)$ representing the circularly symmetric complex Gaussian (CSCG) random variable with mean $\mu$ and variance $\sigma^2$. The instantaneous achievable throughput (in bits/Hz) by Tx$_k$ in $i$th slot is given by Shannon's capacity formula as \cite{info_theory_book}:
\begin{align}
R_{k,i}(P_{k,i})=\tau\log_2\left(1+\frac{|g_{k,i}|^2P_{k,i}}{\sigma_n^2}\right). 
\label{eq:rate}
\end{align}
\subsection{Energy Causality Constraint}
At an EH transmitter, the \textit{energy causality constraint} governs the transmit power. This constraint states that, for Tx$_{k\in \lbrace1, 2, \dotsc, K\rbrace}$, the total energy consumed upto slot $i$ cannot exceed the total energy harvested upto slot $i-1$, plus the initial energy in the battery.
\subsection{Imperfect CSI with Bounded Uncertainty}
Using the pilot symbol based method, the channel coefficient between a transmitter Tx$_k$ and the receiver can be estimated at the receiver. However, due to the noise, the channel coefficient is often estimated erroneously. For Tx$_k$, in $i$th slot, the actual channel coefficient $g_{k,i}$ can be given as
\begin{align}
g_{k,i}=\hat{g}_{k,i}+\Delta g_{k,i},\quad\forall k,i, \label{eq:channel_uncert}
\end{align}
where $\hat{g}_{k,i}$ and $\Delta g_{k,i}$ are the estimated channel coefficient and the estimation error, respectively. We consider a bounded uncertainty model that requires no statistical information about the channel estimation error. Specifically, we bound the estimation error as $|\Delta g_{k,i}|\leq \epsilon$, where $\epsilon>0$ is the radius of the uncertainty region. The estimated channel coefficient $\hat{g}_{k,i}$ is modeled as $\mathcal{CN}(0,1)$. In this case, when $k$th transmitter transmits in $i$th slot, the instantaneous achievable throughput is given as
\begin{align}
R_{k,i}(P_{k,i})=\tau\log_2\left(1+\frac{|\hat{g}_{k,i}+\Delta g_{k,i}|^2P_{k,i}}{\sigma_n^2}\right). \label{eq:rate_k_i_imperfect_CSI}
\end{align}
\textbf{Remark:} Similar to conventional nodes, the channel estimation in an EH network can be done using a standard channel estimation/prediction technique \cite{channel_estimation_survey_2,fading_prediction,pilot_based_prediction_1,channel_estimation_survey}. The extension of the proposed policies incorporating the energy consumed in pilot symbol and energy information transmission could be as follows. Let $P_c$ be the power of the symbols containing pilot symbols and energy information transmitted by each transmitter. In the first slot, each of the $k$th transmitter transmits its pilot and energy information for $\frac{\tau}{K}$ fraction of the slot using TDMA. We assume that $P_c\left(\frac{\tau}{K}\right)\leq E_0^k,\,\forall k$, \textit{i.e.}, the batteries have sufficient initial energy to transmit pilot symbols. Then using the channel estimates and energy information, the receiver obtains the optimal scheduling policy by considering the initial available energy at $k$th transmitter to be $E_0^k-P_c\left(\frac{\tau}{K}\right)$. 
%In our system model, due to the orthogonality between the variables $\mat{W}$ and $\mat{P}$, the receiver broadcasts only the transmit power information to all the transmitters using an ideal backhaul.
% We consider a bounded uncertainty model where the estimation error is a random variable $\Delta g_{k,i}\sim\mathcal{U}[-\epsilon,\epsilon]$ for $\epsilon \geq 0$, where $\mathcal{U}[a,b]$ represents a uniform probability density function between $a$ and $b$
\section{Sum-Rate Maximization with perfect CSI}
\label{sec:sum_rate}
In this section, we formulate the SRM problem under the assumption of perfect CSI at the receiver. The objective is to maximize the sum-rate $R_{\text{sum}}$ of all $K$ transmitters over $M$ slots. Since only one transmitter is allowed to transmit in a slot, the achievable rate in a slot can be characterized using an indicator variable $w_{k,i}$ defined as
\begin{align*}
w_{k,i} = \left\{
\begin{array}{ll}
0, & \text{transmitter $k$ harvests in $i$th slot,}\\
1, & \text{transmitter $k$ transmits in $i$th slot.}
\end{array}
\right.
\end{align*}
Let us consider a $K\times M$ matrix $\mat{W}=[\v{w}_1,\ldots,\v{w}_M]$ where each $K$ dimensional vector $\v{w}_i$ contains the values of $w_{k,i}$ for $i$th slot. Hence, we have $\|\v{w}_i\|_1=1\; \forall i$, \textit{i.e.}, only one transmitter transmits in a slot. The achievable sum-rate in $i$th slot follows as
\begin{align*}
R_i(\v{w}_{i},P_{k,i})=&\sum_{k=1}^K\tau w_{k,i}\log_2\left( 1+\frac{|g_{k,i}|^2P_{k,i}}{\sigma_n^2}\right),\; \forall i,
\end{align*}
and the achievable sum-rate over $M$ slots is given as
\begin{align}
R_{\rm{sum}}(\mat{W},\mat{P})=\sum_{i=1}^M\sum_{k=1}^K\tau w_{k,i}\log_2\left( 1+\frac{|g_{k,i}|^2P_{k,i}}{\sigma_n^2}\right),\label{eq:r_sum}
\end{align}
where $\mat{P}$ is a $K\times M$ matrix whose $(k,i)$th element $[\mat{P}]_{k,i}=P_{k,i}$ represents the power transmitted by $k$th transmitter in $i$th slot. The optimization problem ($\mat{P}_1^{\rm{sum}}$) of maximizing the sum-rate subject to the energy causality constraint is given as
\begingroup
\allowdisplaybreaks
\begin{subequations}
\begin{align}
\mat{P}_1^{\rm{sum}}:\;\max_{\mat{W},\mat{P}}\quad & R_{\rm{sum}}(\mat{W},\mat{P}) \label{eq:max_rate_original_obj}\\
\t{s.t.} \quad & \tau w_{k,1} P_{k,1}\leq E_0^k ,\qquad k=1,\ldots,K, \label{eq:max_rate_original_c1}\\
& \t{(Energy causality constraint for $1$st slot)}\nonumber\\
& \sum_{j=1}^i\tau w_{k,j}P_{k,j}\leq E_0^k+\sum_{j=1}^{i-1}(1-w_{k,j})E_H^{k,j},\nonumber\\
&\qquad\quad\quad k=1,\ldots,K,\,i=2,\ldots,M,\label{eq:max_rate_original_c2}\\
&\hspace{-1cm}\t{(Energy causality constraint for $2$nd to $M$th slot)}\nonumber\\
& \sum_{k=1}^Kw_{k,i}=1,\qquad i=1,\ldots,M, \label{eq:max_rate_original_c3}\\
& (\t{Time sharing constraint})\nonumber\\
& w_{k,i}\in\{0,1\},\;\; P_{k,i}\geq0,\nonumber\\
&\qquad\quad\quad k=1,\ldots,K,\,i=1,\ldots,M, \label{eq:max_rate_original_c4}
\end{align}
\label{eq:max_rate_original}
\end{subequations}
\endgroup
\hspace{-4mm} where $\tau w_{k,i}P_{k,i}$ represents the energy consumed by $k$th transmitter in $i$th slot. Observe that the optimization problem $\mat{P}_1^{\rm sum}$ is a non-convex MINLP due to coupled variables $w_{k,i}$ and $P_{k,i}$. However, we can exploit the binary nature of the variable $w_{k,i}$ to decouple and reduce the problem $\mat{P}_1^{\rm sum}$ to a convex MINLP \cite{cvx_book} $\mat{P}_2^{\rm sum}$ given as
\begingroup
\allowdisplaybreaks
\begin{subequations}
\begin{align}
\mat{P}_2^{\rm{sum}}:\;\max_{\mat{W},\mat{P}}\quad & \sum_{i=1}^M\sum_{k=1}^K\tau \log_2\left(1+\frac{|g_{k,i}|^2P_{k,i}}{\sigma_n^2}\right) \label{eq:max_rate_convex_obj}\\
\t{s.t.} \quad & \tau P_{k,1}\leq w_{k,1}E_0^k ,\qquad k=1,\ldots,K, \label{eq:max_rate_convex_c1}\\
& \tau P_{k,i}\leq w_{k,i}\left[E_0^k+\sum_{j=1}^ME_H^{k,j}\right],\nonumber\\
&\qquad\quad\quad k=1,\ldots,K,i=2,\ldots,M, \label{eq:max_rate_convex_c2}\\
& \sum_{j=1}^i\tau P_{k,j}\leq E_0^k+\sum_{j=1}^{i-1}(1-w_{k,j})E_H^{k,j},\nonumber\\
&\qquad\quad\quad k=1,\ldots,K,i=2,\ldots,M,\label{eq:max_rate_convex_c3}\\
& \sum_{k=1}^Kw_{k,i}=1,\qquad i=1,\ldots,M, \label{eq:max_rate_convex_c4}\\
& w_{k,i}\in\{0,1\},\;\; P_{k,i}\geq0,\nonumber\\
&\qquad\quad\quad k=1,\ldots,K,i=1,\ldots,M. \label{eq:max_rate_convex_c5}
\end{align}
\label{eq:max_rate_convex}
\end{subequations}
\endgroup 
\hspace{-4mm} The correspondence between \eqref{eq:max_rate_original} and \eqref{eq:max_rate_convex} can be understood as follows. When $w_{k,i}=0$ for a fixed $(k,i)$, constraint \eqref{eq:max_rate_convex_c1} or \eqref{eq:max_rate_convex_c2} results in $P_{k,i}\leq 0$, which along with constraint \eqref{eq:max_rate_convex_c5} results in $P_{k,i}=0$. In this case, the constraint \eqref{eq:max_rate_convex_c3} would have no effect as the right-hand side of the inequality is a positive number. On the other hand when $w_{k,i}=1$, the constraint \eqref{eq:max_rate_convex_c2} gives an outer bound on $P_{k,i}$ and hence, has no effect. In this case, the constraint \eqref{eq:max_rate_convex_c3} dominates and represents the energy causality constraint in \eqref{eq:max_rate_original_c2}.

The problem \eqref{eq:max_rate_convex} is a convex MINLP as the objective function is concave in $\mat{P}$ and constraints are linear inequalities in $\mat{P}$ and $\mat{W}$. Since the optimization problem is now linearly separable in variables $\mat{P}$ and $\mat{W}$, it can be efficiently solved using the GBD algorithm \cite{GBD,MINLP_floudas}. In the following subsection, we describe the optimal scheduling policy using the GBD algorithm.

\subsection{Optimal Scheduling Policy using the GBD Algorithm}
\label{sec:optimal_policy_using_GBD}
In \cite{Benders}, an approach to solve mixed integer linear programs (MILPs) with complicating variable was proposed. In these problems, once the complicating variable is fixed, the resulting problem might become tractable and could be parameterized by the value of this complicating variable. Then the optimal value of the complicating variable is obtained using the cutting-plane approach. In \cite{GBD}, the author extended this work for MINLPs and employed a non-linear convex duality theory to obtain the natural families of cuts.

The GBD algorithm \cite{GBD} decomposes the optimization problem $\mat{P}_2^{\rm{sum}}$ into two subproblems: 1) a primal problem (with respect to a real variable) and 2) a master problem (with respect to an integer variable). In our problem, the variable $\mat{W}$ is the complicating variable. Fixing this variable results in a primal problem that is parameterized by the value of $\mat{W}$. In each iteration, the algorithm solves the primal problem and gives a solution $\mat{P}$ along with Lagrange multipliers for fixed $\mat{W}$, which is obtained from the previous iteration of the master problem. Then, for the given solution $\mat{P}$ and the corresponding Lagrangian of current primal problem, the algorithm solves the master problem and obtains $\mat{W}$, which is then passed to the next iteration of the primal problem.  This process is repeated until the convergence is reached. The GBD algorithm is initiated by considering some feasible value of $\mat{W}$ as $\mat{W^{(0)}}$ and solving the first iterate of the primal problem for this $\mat{W^{(0)}}$. The primal and master problems for $l$th iteration are given as follows:
\subsubsection{Primal problem ($l$th iteration)}
\label{subsubsec:primal}
Let $\mat{W}^{*(l-1)}$ be the solution of the master problem in $(l-1)$th iteration. Then the primal problem for the $l$th iteration is given as
\begingroup
\allowdisplaybreaks
\begin{subequations}
\begin{align}
\max_{P_{k,i}\geq0,\, \forall k,i}\quad & \sum_{i=1}^M\sum_{k=1}^K\tau \log_2\left(1+\frac{|g_{k,i}|^2P_{k,i}}{\sigma_n^2}\right) \label{eq:max_rate_primal_obj}\\
\t{s.t.} \quad & \tau P_{k,1}\leq w_{k,1}^{*(l-1)}E_0^k, \qquad k=1,\ldots,K, \label{eq:max_rate_primal_c1}\\
& \tau P_{k,i}\leq w_{k,i}^{*(l-1)}\left[E_0^k+\sum_{j=1}^ME_H^{k,j}\right],\nonumber\\ 
&\qquad\quad\quad k=1,\ldots,K,i=2,\ldots,M,\label{eq:max_rate_primal_c2}\\
& \!\!\!\!\!\sum_{j=1}^i\tau P_{k,j}\leq E_0^k+\sum_{j=1}^{i-1}\left(1-w_{k,j}^{*(l-1)}\right)E_H^{k,j},\nonumber\\
&\qquad\quad\quad k=1,\ldots,K,i=2,\ldots,M,\label{eq:max_rate_primal_c3}
\end{align}
\end{subequations}
\endgroup
where $w_{k,i}^{*(l-1)}$ represents the entry in $\mat{W}^{*(l-1)}$ corresponding to $k$th row and $i$th column.

The primal problem (\ref{eq:max_rate_primal_obj})-(\ref{eq:max_rate_primal_c3}) is a convex optimization problem \cite{cvx_book} in $\mat{P}$. Therefore, an optimal solution can be efficiently obtained using any standard convex optimization solver such as CVX \cite{cvx}. The primal problem can be decomposed and solved in a distributive manner among the users. The solution of the primal problem obtained in $l$th iteration, $\mat{P}^{*(l)}$ is then used to obtain the optimal solution of the master problem in $l$th iteration, $\mat{W}^{*(l)}$. The primal problem is convex with affine constraints, and the constraint set is non-empty. Hence the duality gap is zero and the Karush-Kuhn-Tucker (KKT) stationarity conditions are necessary and sufficient for optimality \cite{cvx_book}. The Lagrangian $\mathcal{L}(\mat{P},\mat{W},\vec{\pmb{\lambda}},\pmb{\gamma},\pmb{\theta})$ of the primal problem is given in \eqref{eq:primal_lagrangian} at the top of the next page, where $\mathcal{X}=\{\mat{P},\mat{W}\}$ and $\mathcal{Y}=\{\vec{\lambda},\pmb{\gamma},\pmb{\theta}\}$ are the sets of primal and dual variables, respectively.
%%%%%%%%%%%%%%%%%%%%%%%%%%%%%%%%%%%%%%%%%%%%%%%%%%%%%%%%%%%%%%%%%%%%%%%%%%%%%%%
\begin{figure*}[!h]
	\begin{multline}
\C{L}(\mathcal{X},\mathcal{Y})=\sum_{k=1}^K\sum_{i=1}^M\tau \log_2\left(1+\frac{|g_{k,i}|^2P_{k,i}}{\sigma_n^2}\right)+\sum_{k=1}^K\lambda_k(w_{k,1}E_0^k-\tau P_{k,1})+ \sum_{k=1}^K\sum_{i=1}^{M-1}\gamma_{k,i}\\ \!\!\!\times \left[w_{k,i+1}\left[E_0^k+\sum_{j=1}^ME_H^{k,j}\right]-\tau P_{k,i+1}\right]
+\sum_{k=1}^K\sum_{i=1}^{M-1}\theta_{k,i}\left[E_0^k+\sum_{j=1}^i(1-w_{k,j})E_H^{k,j}-\sum_{j=1}^{i+1}\tau P_{k,j}\right].
\label{eq:primal_lagrangian}
\end{multline}
\hrulefill
\end{figure*}
%%%%%%%%%%%%%%%%%%%%%%%%%%%%%%%%%%%%%%%%%%%%%%%%%%%%%%%%%%%%%%%%%%%%%%%%%%%%%%%

The KKT stationarity conditions for $k$th transmitter are
\begin{align}
\frac{\tau |g_{k,1}|^2}{\sigma_n^2+|g_{k,1}|^2P^*_{k,1}}-\tau \lambda^*_k-\tau \sum_{j=1}^{M-1}\theta^*_{k,j}&=0, \label{eq:max_rate_KKT_1}\\
\frac{\tau |g_{k,i}|^2}{\sigma_n^2+|g_{k,i}|^2P^*_{k,i}}-\tau \gamma^*_{k,i-1}-\tau \sum_{j=i-1}^{M-1}\theta^*_{k,j}&=0,\nonumber\\ \text{for }i=&2,\ldots,M. \label{eq:max_rate_KKT_i}
\end{align}
The complimentary slackness conditions for $k$th transmitter are
\begingroup
\allowdisplaybreaks
\begin{align}
\lambda^*_k\left(\tau P^*_{k,1}-w_{k,1}E_0^k\right)&=0, \label{eq:CS_max_rate_1}\\
\sum_{i=1}^{M-1}\gamma^*_{k,i}\left[\tau P^*_{k,i+1}-w_{k,i+1}\left[E_0^k+\sum_{j=1}^ME_H^{k,j}\right]\right]&=0,\label{eq:CS_max_rate_2}\\
\sum_{i=1}^{M-1}\theta^*_{k,i}\left[\sum_{j=1}^{i+1}\tau P^*_{k,j}-E_0^k-\sum_{j=1}^i(1-w_{k,j})E_H^{k,j}\right]&=0,\label{eq:CS_max_rate_3}
\end{align}
\endgroup
where $\vec{\pmb{\lambda}}\in\mathbb{R}_+^{K\times1}$, $\pmb{\gamma}\in\mathbb{R}_+^{K\times (M-1)}$, and $\pmb{\theta}\in\mathbb{R}_+^{K\times (M-1)}$ are dual variables associated with constraints (\ref{eq:max_rate_primal_c1}), (\ref{eq:max_rate_primal_c2}), and (\ref{eq:max_rate_primal_c3}), respectively. For simplicity, we omit the non-negativity constraint on $\mat{P}$, which can be incorporated later by projecting the optimal solution onto the positive orthant. Using the KKT conditions, the optimal transmit power for $k$th transmitter in $l$th iteration is given as
\begingroup
\allowdisplaybreaks
\begin{align}
P^*_{k,1}&=\left[\frac{1}{\lambda^*_k+\sum_{i=1}^{M-1}\theta^*_{k,j}}-\frac{\sigma_n^2}{|g_{k,1}|^2}\right]^+,\label{eq:max_rate_power_1}\\
P^*_{k,i}&=\left[\frac{1}{\gamma^*_{k,i-1}+\sum_{j=i-1}^{M-1}\theta^*_{k,j}}-\frac{\sigma_n^2}{|g_{k,i}|^2}\right]^+,\nonumber\\
&\qquad\qquad\qquad\qquad\qquad\qquad \text{for }i=2,\ldots,M,\label{eq:max_rate_power_2}
\end{align}
\endgroup
where $x^+=\max\{0,x\}$ represents the projection onto the positive orthant. Equations \eqref{eq:max_rate_power_1} and \eqref{eq:max_rate_power_2} require the optimal values of dual variables $\vec{{\lambda}}^*$, $\pmb{\gamma}^*$ and $\pmb{\theta}^*$, which are obtained using CVX \cite{cvx}. Alternatively, these values can be obtained using the iterative dual-descent method \cite{dual_descent,bedi_tcom}. Since the objective function of the primal problem is concave and the inequalities are linear, the duality gap is zero.
\subsubsection{Master problem ($l$th iteration)}
The master problem from the original optimization problem \eqref{eq:max_rate_convex} is obtained using the following two manipulations \cite{GBD}:
\begin{enumerate}
	\item Projecting (6) onto $\mat{W}$-space as
	\begingroup
	\allowdisplaybreaks
	\begin{align*}
	\max_{\mat{W}\in\mathcal{W}} \quad & v(\mat{W})
	\end{align*}
	\endgroup
	where 
	\begin{align*}
	v(\mathbf{W})=\left\{ \begin{array}{ll}
	\sup\limits_{\mathbf{P}}\quad & \sum\limits_{i=1}^M\sum\limits_{k=1}^K\tau \log_2\left(1+\frac{|g_{k,i}|^2P_{k,i}}{\sigma_n^2}\right)\vspace{2mm}\\ \vspace{2mm}
	\text{s.t.} \quad & {\rm (6b)-(6d)} ,\; P_{k,i}\geq 0 ,\quad \forall k,i.
	\end{array}\right.
	\end{align*}
	Note that $v(\mat{W})$ is the primal problem discussed above.
	\item Summoning the natural dual representation of $v$ in terms of the pointwise infimum of a collection of functions that dominates it.
\end{enumerate}
These manipulations result in the master problem for the $l$th iteration given as \cite{GBD}
\begingroup
\allowdisplaybreaks
\begin{subequations}
	\label{eq:max_master}
	\begin{align}
	\max_{\mat{W},\beta} \quad & \beta \label{eq:max_master_obj}\\
	\t{s.t.} \quad & \beta\leq \C{L}(\mat{P}^{*(j)},\vec{\lambda}^{*(j)},\pmb{\gamma}^{*(j)},\pmb{\theta}^{*(j)}),\; j\in\{1,2,\ldots,l\}, \label{eq:max_master_c1}\\
	& \mat{W}\in\{0,1\}^{K\times M} ,\quad \sum_{k=1}^K w_{k,i}=1,\; \forall i, \label{eq:max_master_c2}\\
	& \beta\geq0, \label{eq:max_master_c3}
	\end{align}
\end{subequations}
\endgroup
where $\C{L}(\mat{P}^{*(l)},\vec{\lambda}^{*(l)},\pmb{\gamma}^{*(l)},\pmb{\theta}^{*(l)})$ is the Lagrangian of the primal problem and $\{\vec{\lambda}^{*(l)},\pmb{\gamma}^{*(l)},\pmb{\theta}^{*(l)}\}$ is the set of optimal dual variables corresponding to the constraints \eqref{eq:max_rate_primal_c1}, \eqref{eq:max_rate_primal_c2}, and \eqref{eq:max_rate_primal_c3}, respectively, obtained by solving the $l$th iteration of the primal problem.

The problem \eqref{eq:max_master} is a mixed integer linear program (MILP) of $\beta$ and $\mat{W}$. Therefore, an optimal solution can be efficiently obtained using any standard MILP solver, \textit{e.g.}, MOSEK \cite{mosek}. 

\textbf{Generalized Benders decomposition algorithm:} The master problem gives a solution $\beta^{*(l)}$ in the $l$th iteration. This $\beta^{*(l)}$ upper bounds the optimal solution of the original problem $\mat{P}_2^{\rm{sum}}$. In addition, in each iteration, an extra constraint \eqref{eq:max_master_c1} is being added to the master problem. Hence the upper bound $\beta^{*(l)}$ is non-increasing with the number of iterations.

The solution of the primal problem in $l$th iteration, $\mat{P}^{*(l)}$ lower bounds the optimal solution of $\mat{P}_2^{\rm{sum}}$ by solving $\mat{P}_2^{\rm{sum}}$ for a fixed $\mat{W}$, \textit{i.e.}, $\mat{W}^{*(l-1)}$. The lower bound in each iteration is set to be the maximum of the lower bounds obtained until the current iteration.

In the $l$th iteration, the primal problem is solved for the solution obtained by the master problem in $(l-1)$th iteration. Then, for the solution obtained by the primal problem in $l$th iteration, we solve the $l$th iteration of the master problem. This process continues, and due to non-increasing (non-decreasing) nature of the upper (lower) bound, the GBD algorithm converges to the optimal solution in a finite number of iterations \cite{GBD,MINLP_floudas}. The GBD algorithm is summarized in Algorithm \ref{algo:GBD}, where $\mathcal{S}$ is a set of constraint (\ref{eq:max_master_c1}) in which an additional constraint is added in each iteration.
\begin{algorithm}
\caption{GBD algorithm}
\label{algo:GBD}
\begin{algorithmic}
\State \textbf{Initialization:} Initialize $\mat{W}^{(0)}$ and convergence parameter $\zeta$. Set $\mathcal{S}\leftarrow\emptyset$ and $j\leftarrow1$.
\State Set $\text{flag}\leftarrow1$
\While{$\text{flag}\neq0$}
\State Solve the primal problem (\ref{eq:max_rate_primal_obj})-(\ref{eq:max_rate_primal_c3}) and obtain
\{$\mat{P}^{*},\bar{\pmb{\lambda}}^*,\pmb{\gamma}^*,\pmb{\theta}^*$\} and lower bound L$_{(j)}$
\State $\mathcal{S}\leftarrow \mathcal{S}\cup\{j\}$
\State Solve master problem (\ref{eq:max_master_obj})-(\ref{eq:max_master_c3}) and obtain $\mat{W}^{(j)*}$ and the upper bound U$_{(j)}$.
\If{$\vert$U$_{(j)}$-L$_{(j)}\vert\leq\zeta$}
\State $\text{flag}\leftarrow0$
\EndIf
\State Set $j\leftarrow j+1$
\EndWhile \\
\Return $\mat{P}$ and $\mat{W}$
\end{algorithmic}
\end{algorithm}
The primal problem is convex and hence can be efficiently solved in polynomial time. The master problem on the other hand is an MILP and hence has non-polynomial complexity. However, the master problem can be efficiently solved using MOSEK \cite{mosek} as the GBD algorithm is executed offline.\footnote{The channel coefficients and energy informations are assumed to be known at the receiver in advance. Thus the SRM and MRM scheduling problems fall in the category of \textit{offline optimization} framework, which can be studied using the GBD algorithm.} We below provide the proof that the GBD algorithm converges in a finite number of iterations.
\begin{theorem*}
The GBD algorithm for MINLP $\mat{P}_2^{\rm{sum}}$ converges to a $\zeta$-optimal solution in a finite number of iterations for any $\zeta\geq0$.
\end{theorem*}
\begin{IEEEproof}
The GBD algorithm achieves a $\zeta$-optimal solution if $\left|\rm{U}_{(j)}-\rm{L}_{(j)}\right|\leq \zeta$, for any $\zeta\geq0$ where $j$ is the iteration number \cite{GBD,MINLP_floudas}. Let $\mathcal{P}\subseteq \mathbb{R}_+^{K\times M}$ and $\mathcal{W}=\{0,1\}^{K\times M}$ be the sets such that
\begingroup
\allowdisplaybreaks
\begin{align*}
\mathcal{P}&=\{\mat{P}:P_{k,i}\geq0,\, \forall k,i\} \subseteq \mathbb{R}_+^{K\times M} \\ 
\mathcal{W}&=\{\mat{W}:w_{k,i}\in\{0,1\},\, \forall k,i\},
\end{align*}
\endgroup
and $\vec{g}(\mat{P},\mat{W}):\mathcal{P}\times\mathcal{W}\rightarrow\mathcal{X}\subseteq \mathbb{R}^p,\,\vec{h}(\mat{P},\mat{W}):\mathcal{P}\times\mathcal{W}\rightarrow\mathcal{Y}\subseteq \mathbb{R}^q$ be the functions such that $\vec{g}(\mat{P},\mat{W})\preceq\vec{0}$ corresponds to the constraints (\ref{eq:max_rate_convex_c1})-(\ref{eq:max_rate_convex_c3}) and $\vec{h}(\mat{P},\mat{W})=\vec{0}$ corresponds to the constraint (\ref{eq:max_rate_convex_c4}), where $p$ and $q$ represent the number of inequality and equality constraints, respectively. 

In $\mat{P}_2^{\rm{sum}}$, the set $\mathcal{P}$ is a non-empty convex set and the functions $\vec{g}(\mat{P},\mat{W})$ and $\vec{h}(\mat{P},\mat{W})$ are convex and affine, respectively. Also, both functions are continuous for each fixed $\mat{W}\in\mathcal{W}=\{0,1\}^{K\times M}$. In addition, for each $\mat{W}\in\mathcal{W}$, the problem $\mat{P}_2^{\rm{sum}}$ is a convex optimization problem \cite{cvx_book} in $\mat{P}\in\mathcal{P}$ and, has a finite optimal solution ($\mat{P}^*$) and optimal Lagrange multiplier vectors ($\bar{\pmb{\lambda}}^*,\pmb{\gamma}^*,\pmb{\theta}^*$) for inequalities and equalities. Therefore, as per the steps outlined in \cite{GBD}, the convergence holds for the problem $\mat{P}_2^{\rm{sum}}$ for any $\zeta\geq0$.
\end{IEEEproof}

As stated earlier, the optimization problem in (\ref{eq:max_rate_convex}) cannot be solved in polynomial time. In the next subsection, we present a low-complexity suboptimal scheduling policy, which can be efficiently solved in polynomial time.
\subsection{Suboptimal Scheduling Policy}
\label{sec:suboptimal_policy}
In this subsection, we propose a low-complexity suboptimal scheduling policy which can be obtained in polynomial time. To obtain a low-complexity suboptimal scheduling policy, we first replace the non-convex set $\mathcal{NC}=\{w_{k,i}: w_{k,i}\in\{0,1\},\;\forall k,i\}$ with a convex relaxation $\mathcal{C}=\{w_{k,i}:0\leq w_{k,i}\leq 1,\; \forall k,i\}$ such that $\mathcal{NC}\subset \mathcal{C}$, and then formulate a relaxed version of the optimization problem $\mat{P}_2^{\rm{sum}}$, $\mat{P}^{\rm{rel}}$ as
\begingroup
\allowdisplaybreaks
\begin{subequations}
\label{eq:sum_rate_relaxed}
\begin{align}
\mat{P}^{\rm{rel}}: \; \max_{\mat{W},\mat{P}} \quad & \sum_{i=1}^M\sum_{k=1}^K\tau \log_2\left(1+\frac{|g_{k,i}|^2P_{k,i}}{\sigma_n^2}\right) \label{eq:relaxed_obj}\\
\t{s.t.} \quad & \eqref{eq:max_rate_convex_c1}-\eqref{eq:max_rate_convex_c4}, \label{eq:relaxed_c1}\\
& P_{k,i}\geq0,\quad w_{k,i}\geq0,\;\; \forall k,i. \label{eq:relaxed_c2}
\end{align}
\end{subequations}
\endgroup

Note that we have omitted the constraint $w_{k,i}\leq1,$ which can be incorporated through the constraint $\sum_{k=1}^Kw_{k,i}=1,\; \forall i$. The solution $(\mat{W}^*_{\rm{rel}},\mat{P}^*_{\rm{rel}})$ of this relaxed problem $\mat{P}^{\rm{rel}}$ upper bounds the solutions of \eqref{eq:max_rate_original} and \eqref{eq:max_rate_convex} as this solution belongs to the set $\C{C}$ and can be infeasible for both the problems \eqref{eq:max_rate_original} and \eqref{eq:max_rate_convex}. Hence, we need to project them onto the feasible set $\C{NC}$. For this, we first round off the variable $\mat{W}^*_{\rm{rel}}$ to the nearest integer as $\mat{W}^*_{\rm{sub}}=\rm{round}(\mat{W}^*_{\rm{rel}})$ such that $\mat{W}^*_{\rm{sub}}\in\C{NC}$ and then obtain $\mat{P}^*_{\rm{sub}}$ by solving \eqref{eq:max_rate_convex} for fixed $\mat{W}^*_{\rm{sub}}$. The suboptimal scheduling algorithm is given in Algorithm \ref{algo:suboptimal}.
\begin{algorithm}
	\caption{Suboptimal algorithm}
	\label{algo:suboptimal}
	\begin{algorithmic}
		\State \textbf{Initialization:} Solve Relaxed problem $\mat{P}^{\rm{rel}}$ and obtain $\mat{W}^*_{\rm{rel}}$.
		\State \textbf{Approximate:} $\mat{W}^*_{\rm{sub}}:=\t{round}(\mat{W}^*_{\rm{rel}})$.
		\State Solve the optimization problem (\ref{eq:max_rate_convex}) for a fixed $\mat{W}^*_{\rm{sub}}$ and obtain $\mat{P}^*_{\rm{sub}}$.\\
		\Return $\mat{P}^*_{\rm{sub}}$ and $\mat{W}^*_{\rm{sub}}$
	\end{algorithmic}
\end{algorithm}

From the optimal scheduling policy obtained for the SRM problem, it can be noted that a transmitter closer to the receiver achieves a higher rate than that of the farther one. This is because the channel power gain $h_k$ is proportional to $D_k^{-\alpha}$, where $\alpha\geq2$ denotes the path loss exponent and $D_k$ denotes the distance between $k$th transmitter and the receiver. This results in an unfair rate allocation among transmitters. One way to tackle such an unfair rate allocation issue is to maximize the minimum achievable rate. We discuss the problem of the MRM in Section \ref{sec:min_rate}.
\section{Minimum-Rate Maximization with perfect CSI}
\label{sec:min_rate}
In the MRM problem, the goal is to maximize the minimum rate in the network. The optimization problem $\mat{P}_1^{\rm{min}}$ for this policy is given as
\begingroup
\allowdisplaybreaks
\begin{subequations}
\begin{align}
\max_{\mat{W},\mat{P},\bar{R}} \quad & \bar{R} \label{eq:min_rate_orig_obj}\\
\t{s.t.} \quad & \min_k\left\{\sum_{i=1}^M \tau w_{k,i} \log_2\left(1+\frac{|g_{k,i}|^2P_{k,i}}{\sigma_n^2}\right)\right\}\geq \bar{R}, \label{eq:min_rate_orig_c1}\\
& \eqref{eq:max_rate_original_c1}-\eqref{eq:max_rate_original_c4},\label{eq:min_rate_orig_c2}
\end{align}
\end{subequations}
\endgroup
where $\bar{R}$ is the target rate. The problem (\ref{eq:min_rate_orig_obj})-(\ref{eq:min_rate_orig_c2}) is also a non-convex MINLP problem in $\mat{W},\mat{P}$ and $\bar{R}$, and can be converted into a convex MINLP as discussed before in Section~\ref{sec:sum_rate} and solved efficiently using the GBD algorithm. The convex MINLP $\mat{P}_2^{\rm{min}}$ can be given as
\begin{subequations}
\begin{align}
\max_{\mat{W},\mat{P},\bar{R}} \quad & \bar{R} \label{eq:min_rate_convex_obj}\\
\t{s.t.} \quad & \min_k\left\{\sum_{i=1}^M \tau \log_2\left(1+\frac{|g_{k,i}|^2P_{k,i}}{\sigma_n^2}\right)\right\}\geq \bar{R}, \label{eq:min_rate_convex_c1}\\
& \eqref{eq:max_rate_convex_c1}-\eqref{eq:max_rate_convex_c5}.\label{eq:min_rate_convex_c2}
\end{align}
\end{subequations}
The primal and the master problems for the MINLP (\ref{eq:min_rate_convex_obj})-(\ref{eq:min_rate_convex_c2}) are given in the following subsections.
\subsection{Primal Problem ($l$th iteration)}
The primal problem in the $l$th iteration is given as
\begin{subequations}
\begin{align}
\max_{\mat{P},\bar{R}}\quad & \bar{R} \label{eq:primal_min_rate_obj}\\
\t{s.t.}\quad & \eqref{eq:min_rate_convex_c1} \t{ and } \eqref{eq:max_rate_primal_c1}-\eqref{eq:max_rate_primal_c3}, \label{eq:primal_min_rate_c1}
\end{align}
\end{subequations}
which is a convex optimization problem in $\mat{P}$ and $\bar{R}$. Hence an optimal solution can be obtained using CVX \cite{cvx} as discussed in Section \ref{subsubsec:primal}. The Lagrangian of the primal problem is given in \eqref{eq:primal_lagrangian_min_rate} at the top of the next page where $\mathcal{X}=\{\bar{R},\mat{P},\mat{W}\}$ and $\mathcal{Y}=\{\vec{\delta},\vec{\lambda},\pmb{\gamma},\pmb{\theta}\}$ are the sets of primal and dual variables, respectively.
%%%%%%%%%%%%%%%%%%%%%%%%%%%%%%%%%%%%%%%%%%%%%%%%%%%%%%%%%%%%
\begin{figure*}[!h]
\begin{multline}
\C{L}(\mathcal{X},\mathcal{Y})=\bar{R}+\sum_{k=1}^K\delta_k\left[\sum_{i=1}^M\tau \log_2\left(1+\frac{|g_{k,i}|^2P_{k,i}}{\sigma_n^2}\right)-\bar{R}\right]+\sum_{k=1}^K\lambda_k(w_{k,1}E_0^k-\tau P_{k,1})\\+\sum_{k=1}^K\!\!\sum_{i=1}^{M-1}\!\gamma_{k,i}\!\left[w_{k,i+1}\!\left[E_0^k\!+\!\sum_{j=1}^M\!\!E_H^{k,j}\right]\!-\!\tau P_{k,i+1}\right]\!+\!\sum_{k=1}^K\!\!\sum_{i=1}^{M-1}\!\theta_{k,i}\!\left[E_0^k\!+\!\sum_{j=1}^i(1-w_{k,j})E_H^{k,j}\!-\!\sum_{j=1}^{i+1}\tau P_{k,j}\right].
\label{eq:primal_lagrangian_min_rate}
\end{multline}
\hrulefill
\end{figure*}
%%%%%%%%%%%%%%%%%%%%%%%%%%%%%%%%%%%%%%%%%%%%%%%%%%%%%%%%%%%%

The KKT stationarity conditions are 
\begin{align}
1-\sum_{k=1}^K\delta_k^*&=0, \label{eq:KKT_min_rate_1}\\
\frac{\tau \delta_k^*|g_{k,1}|^2}{\sigma_n^2+|g_{k,1}|^2P^*_{k,1}}-\tau \lambda_k^*-\sum_{j=1}^{M-1}\tau \theta^*_{k,j}&=0,\nonumber\\ \t{for } k&=1,\ldots,K, \label{eq:KKT_min_rate_2}\\
\frac{\tau \delta_k^*|g_{k,i}|^2}{\sigma_n^2+|g_{k,i}|^2P^*_{k,i}}-\tau \gamma_{k,i-1}^*-\sum_{j=i-1}^{M-1}\tau \theta^*_{k,j}&=0,\nonumber\\
 \t{for } i=2,\ldots,M,\;k&=1,\ldots,K. \label{eq:KKT_min_rate_3}
\end{align}
The complimentary slackness conditions are
\begin{align}
\delta_k^*\left[\bar{R}-\sum_{i=1}^M\tau \log_2\left(1+\frac{|g_{k,i}|^2P_{k,i}}{\sigma_n^2}\right)\right]=0, \label{eq:CS_min_rate_1}\\
\eqref{eq:CS_max_rate_1}-\eqref{eq:CS_max_rate_3}. \label{eq:CS_min_rate_2}
\end{align}
where $\vec{\pmb{\delta}}\in\mathbb{R}_+^{K\times 1}$, $\vec{\pmb{\lambda}}\in\mathbb{R}_+^{K\times 1}$, $\pmb{\gamma}\in\mathbb{R}_+^{K\times (M-1)}$, and $\pmb{\theta}\in\mathbb{R}_+^{K\times (M-1)}$ are dual variables associated with constraints (\ref{eq:min_rate_orig_c1}), (\ref{eq:max_rate_primal_c1}), (\ref{eq:max_rate_primal_c2}), and (\ref{eq:max_rate_primal_c3}), respectively. Using the KKT conditions, the optimal transmit power for $k$th transmitter in the $l$th iteration is given as
\begin{align}
P_{k,1}^*&=\left[\frac{\delta_k^*}{\lambda_k^*+\sum_{j=1}^{M-1}\theta_{k,j}^*}-\frac{\sigma_n^2}{|g_{k,1}|^2}\right]^+, \label{eq:min_rate_power_1}\\
P_{k,i}^*&=\left[\frac{\delta_k^*}{\gamma_{k,i-1}^*+\sum_{j=i-1}^{M-1}\theta_{k,j}^*}-\frac{\sigma_n^2}{|g_{k,i}|^2}\right]^+,\; \forall i\backslash\{1\}. \label{eq:min_rate_power_i}
\end{align}
\vspace{-4mm}
\subsection{Master Problem ($l$th iteration)}
The Lagrangian of the primal problem is given in (\ref{eq:primal_lagrangian_min_rate}). The master problem in $l$th iteration is given as
\begin{subequations}
\begin{align}
\max_{\mat{W},\beta} \quad & \beta \label{eq:master_min_rate_obj}\\
\t{s.t.} \quad & \beta\leq\C{L}\left(\bar{R}^{*(j)},\mat{P}^{*(j)},\vec{\delta}^{*(j)},\vec{\lambda}^{*(j)},\pmb{\gamma}^{*(j)},\pmb{\theta}^{*(j)}\right),\nonumber\\
&\qquad\qquad\qquad\qquad\qquad\quad j\in\{1,2,\ldots,l\} \label{eq:master_min_rate_c1}\\
& \mat{W}\in\{0,1\}^{K\times M}, \quad \sum_{k=1}^K w_{k,i}=1,\,\forall i \label{eq:master_min_rate_c2}\\
& \beta\geq0. \label{eq:master_min_rate_c3}
\end{align}
\end{subequations}
\section{Sum-Rate Maximization with imperfect CSI}
\label{sec:sum_rate_imp}
Under the case of imperfect CSI, we aim to obtain a robust scheduling policy maximizing the worst-case sum-rate of all transmitters. Under the assumption of bounded uncertainty given in (\ref{eq:channel_uncert}), the achievable rate of the $k$th transmitter in $i$th slot in the worst case scenario is given as
\begin{align*}
\!\!\!\!R_{k,i}^{\rm{worst}}\!\!=\!\!\min_{|\Delta g_{k,i}|\leq \epsilon}\!\! \tau  w_{k,i}\log_2\!\!\left(\!1+\frac{|\hat{g}_{k,i}+\Delta g_{k,i}|^2P_{k,i}}{\sigma_n^2}\!\right) .
\end{align*}
Then the optimization problem of maximizing the worst-case achievable sum-rate is given as
\begin{subequations}
\begin{align}
\max_{\mat{W},\mat{P}} \quad & \sum_{i=1}^M\sum_{k=1}^K\tau  w_{k,i}  \min_{|\Delta g_{k,i}|\leq \epsilon} \log_2\left(1+\frac{|\hat{g}_{k,i}+\Delta g_{k,i}|^2P_{k,i}}{\sigma_n^2}\right) \label{eq:max_rate_iCSI_obj_orig}\\
\t{s.t.} \quad & \eqref{eq:max_rate_original_c1}-\eqref{eq:max_rate_original_c4}. \label{eq:max_rate_iCSI_c1_orig}
\end{align}
\end{subequations}
Using the inequality $|\hat{g}_{k,i}+\Delta g_{k,i}|^2\geq |\hat{g}_{k,i}|^2+\epsilon^2-2|\hat{g}_{k,i}|\epsilon$, the objective can further be simplified as
\begingroup
\allowdisplaybreaks
\begin{multline*}
\min_{|\Delta g_{k,i}|\leq \epsilon} \log_2\left(1+\frac{|\hat{g}_{k,i}+\Delta g_{k,i}|^2P_{k,i}}{\sigma_n^2}\right)=\\
\underbrace{\log_2\left(1+\frac{(|\hat{g}_{k,i}|^2+\epsilon^2-2|\hat{g}_{k,i}|\epsilon)P_{k,i}}{\sigma_n^2}\right)}_{\hat{R}_{k,i}^{\rm{worst}}}.
\end{multline*}
\endgroup
The optimization problem is now given as
\begin{subequations}
\begin{align}
\max_{\mat{W},\mat{P}} \quad & \sum_{i=1}^M\sum_{k=1}^K\tau  w_{k,i} \log_2\left(1+\frac{(|\hat{g}_{k,i}|^2+\epsilon^2-2|\hat{g}_{k,i}|\epsilon)P_{k,i}}{\sigma_n^2}\right)
\label{eq:max_rate_iCSI_obj}\\
\t{s.t.} \quad & \eqref{eq:max_rate_original_c1}-\eqref{eq:max_rate_original_c4}. \label{eq:max_rate_iCSI_c1}
\end{align}
\end{subequations}
The problem (\ref{eq:max_rate_iCSI_obj})-(\ref{eq:max_rate_iCSI_c1}) is also an MINLP problem, and an optimal solution can be obtained using the GBD algorithm as discussed in Section~\ref{sec:sum_rate}.
\section{Minimum-Rate Maximization with imperfect CSI}
\label{sec:min_rate_imp}
The issue of unfair rate allocation still persists with the SRM with imperfect CSI. We address this issue by maximizing the minimum worst-case achievable rate in the network subject to energy causality constraints. The corresponding optimization problem is given as
\begingroup
\allowdisplaybreaks
\begin{subequations}
\begin{align}
\max_{\bar{R},\mat{W},\mat{P}} \quad & \bar{R} \label{eq:min_rate_iCSI_obj_orig}\\
\t{s.t.\hspace{2mm}}\quad & \min_k\left\{\sum_{i=1}^M R_{k,i}^{\rm{worst}} \right\}\geq \bar{R}, \label{eq:min_rate_iCSI_c1_orig}\\
& \eqref{eq:max_rate_original_c1}-\eqref{eq:max_rate_original_c4}. \label{eq:min_rate_iCSI_c2_orig}
\end{align}
\end{subequations}
\endgroup
\noindent Similar to (\ref{eq:max_rate_iCSI_obj_orig})-(\ref{eq:max_rate_iCSI_c1_orig}), the optimization problem (\ref{eq:min_rate_iCSI_obj_orig})-(\ref{eq:min_rate_iCSI_c2_orig}) can further be simplified as
\begingroup
\allowdisplaybreaks
\begin{subequations}
\begin{align}
\max_{\bar{R},\mat{W},\mat{P}} \quad & \bar{R} \label{eq:min_rate_iCSI_obj}\\
\t{s.t.\hspace{2mm}}\quad & \min_k\left\{\sum_{i=1}^M \tau w_{k,i}\hat{R}_{k,i}^{\rm{worst}} \right\} \geq \bar{R}, \label{eq:min_rate_iCSI_c1}\\
& \eqref{eq:max_rate_original_c1}-\eqref{eq:max_rate_original_c4}. \label{eq:min_rate_iCSI_c2}
\end{align}
\end{subequations}
\endgroup
which can be solved efficiently using the GBD algorithm as discussed in Section~\ref{sec:sum_rate}.
\section{Myopic Policies}
\label{sec:myopic}
The myopic scheduling policies presented in \cite{TDMA_zhang} and \cite{TDMA_full_duplex_kang} can be modified so that they can be applied to our system model and compared with our proposed policies. We now discuss the modification in the myopic policies.
\begin{enumerate}
	\item \textbf{Myopic policy of \cite{TDMA_zhang}:} The $i$th slot of $k$th transmitter under the scheduling policy of \cite{TDMA_zhang} is shown in Fig. \ref{fig:zhang_slot_paper}.
	\begin{figure}[!ht]
		\centering
		\includegraphics[width=0.9\linewidth]{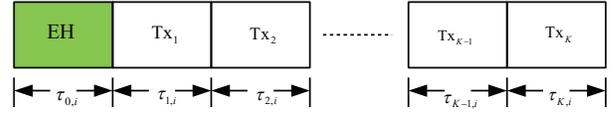}
		\caption{The $i$th slot of $k$th transmitter in myopic policy \cite{TDMA_zhang}.}
		\label{fig:zhang_slot_paper}
	\end{figure}

	Let $E_h^{k,i}$ represent the energy available for harvesting in $i$th slot for $k$th transmitter. If the slot length is assumed to be $T$ seconds, then according to the policy in \cite{TDMA_zhang}, $\tau_{0,i}$ fraction of the slot is reserved for energy harvesting and $\tau_{k,i}$, $k=1,\ldots,K$ fraction of the slot is reserved for data transmission for the $k$th user in $i$th slot. In this case we have
	\begin{align*}
	\sum_{k=0}^K \tau_{k,i}=T,\quad \forall i=1,\ldots,M.
	\end{align*}
	Thus the energy harvested by $k$th transmitter in $i$th slot is $
	E_{\rm Harv}^{k,i}=\left(\frac{\tau_{0,i}}{T}\right) E_h^{k,i}.
$ Without loss of generality, we assume $T=1$ second. Then the transmit power of $k$th transmitter in $i$th slot, $P_{k,i}$ is given as
	\begin{align}
	P_{k,i}=\left\{
	\begin{array}{ll}
	\frac{\tau_{0,i}}{\tau_{k,i}}\cdot (E_h^{k,i}+E_0^k), & \text{for } i=1,\\
	\frac{\tau_{0,i}}{\tau_{k,i}}\cdot E_h^{k,i}, & \text{for } i>1,
	\end{array}
	\right.
	\label{eq:zhang_power_paper}
	\end{align}
	where $E_0^k$ is the initial energy available in the battery of the $k$th transmitter. The aim here is to optimize the time sharing parameter $\tau_{k,i}$ such that the instantaneous sum-rate and minimum-rate of the network are maximized.
	\item \textbf{Myopic policy of \cite{TDMA_full_duplex_kang}:} The $i$th slot of $k$th transmitter under the scheduling policy of \cite{TDMA_full_duplex_kang} is shown in Fig. \ref{fig:full_duplex_slots_paper}.
	\begin{figure}[!ht]
		\centering
		\includegraphics[width=0.9\linewidth]{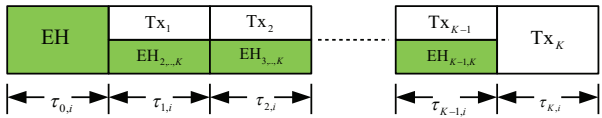}
		\caption{The $i$th slot of $k$th transmitter in myopic policy \cite{TDMA_full_duplex_kang}.}
		\label{fig:full_duplex_slots_paper}
	\end{figure}

	If $E_h^{k,i}$ represent the energy available for harvesting in $i$th slot for $k$th transmitter, then the total energy harvested by the transmitter is $E_{\rm Harv}^{k,i}=\left(\sum_{j=1}^{k-1} \tau_{j,i}\right)  E_h^{k,i}$,	where we have assumed the slot length to be 1 second. Then the transmit power $P^{\rm FD}_{k,i}$ of the $k$th transmitter in $i$th slot is given as
	\begin{align}
	\!\!\!P^{\rm FD}_{k,i}=\left\{\!\!
	\begin{array}{ll}
	\frac{1}{\tau_{k,i}}\cdot \left[\left(\sum\limits_{j=0}^{k-1}\tau_{j,i}\right)E_h^{k,i}+E_0^k\right], & \text{for } i=1,\\
	& \\
	\frac{1}{\tau_{k,i}}\cdot \left[\left(\sum\limits_{j=0}^{k-1}\tau_{j,i}\right)E_h^{k,i}\right], & \text{for } i>1.
	\end{array}
	\right.
	\label{eq:fd_power_paper}
	\end{align}
	The aim here is to obtain an optimal time sharing among the transmitters, $\tau_{k,i}$ such that the instantaneous sum-rate and minimum-rate of the network are maximized.
\end{enumerate}

\section{Results and Discussions}
\label{sec:results}
In this section, we present simulation results for scheduling policies discussed in previous sections. All channel links are subject to independent Rayleigh fading. We assume that all transmitters have an initial energy of $E_0^k=2\t{ mJ},\, k=1,\ldots,K$. The noise power at the receiver is $-30$ dBm. Unless otherwise stated, to gain insights, we focus on the case of two transmitters ($K = 2$), where the distances of transmitters Tx$_1$ and Tx$_2$ from the receiver are 5 and 10 meters, respectively. For the simulation purpose, the harvested energies for both the transmitters are generated uniformly at random, \textit{i.e.}, $\mathcal{U}[0,5]$ mJ where $\mathcal{U}[a,b]$ represents a uniform probability density function between $a$ and $b$. The slot length $\tau$ is assumed to be 1 second.

\subsection{Optimal Policy with Perfect CSI}
\subsubsection{Rate versus number of slots ($M$)}
\begin{figure}[!h]
		\centering
		\includegraphics[width=0.9\linewidth]{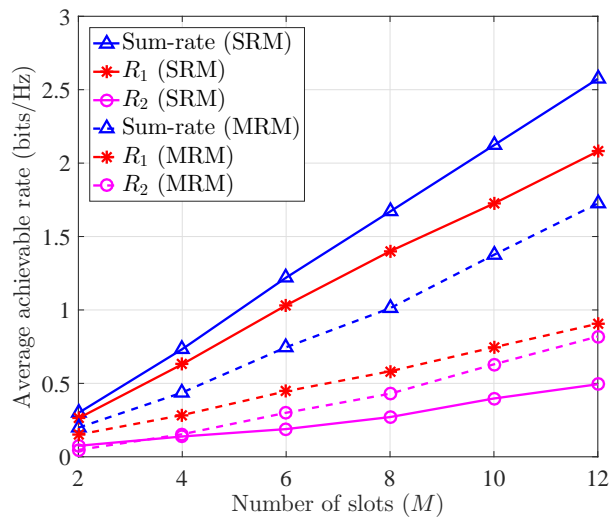}
		\caption{Comparison of achievable rates of transmitters between the SRM and MRM policies.}
		\label{fig:comparison_max_rate_min_rate}
\end{figure}
\begin{figure}[!h]
	\centering
\includegraphics[width=0.9\linewidth]{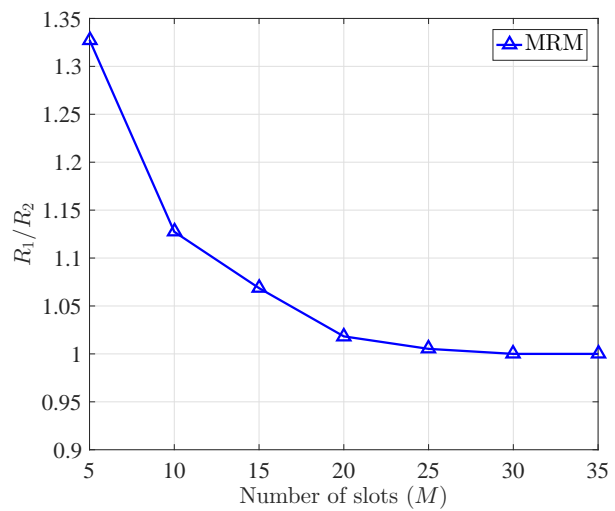}
\caption{Ratio of average achievable rates, $R_1/R_2$ versus the total number of slots ($M$) for the MRM policy with perfect CSI.}
\label{fig:R_1_R_2_ratio_vs_slots}
\end{figure}
%%%%%%%%%%%%%%%%%%%%%%%%%%%%%%%%%%%%%%%%%%%%%%%%%%%%%%%%%%%%%%%%%%%%%%%%%%%%%%%%%%%%%%%
%\begin{figure*}[h]
%\centering
%\begin{minipage}{0.48\linewidth}
%\centering
%\includegraphics[width=0.9\linewidth]{comparison_max_rate_min_rate}
%\caption{Comparison of achievable rates of transmitters between the SRM and MRM policies.}
%\label{fig:comparison_max_rate_min_rate}
%\end{minipage}
%\hspace{3mm}
%\begin{minipage}{0.48\linewidth}
%	\centering
%\includegraphics[width=0.9\linewidth]{R_1_R_2_ratio_vs_slots}
%\caption{Ratio of average achievable rates, $R_1/R_2$ versus the total number of slots ($M$) for the MRM policy with perfect CSI.}
%\label{fig:R_1_R_2_ratio_vs_slots}
%\end{minipage}
%\end{figure*}

%%%%%%%%%%%%%%%%%%%%%%%%%%%%%%%%%%%%%%%%%%%%%%%%%%%%%%%%%%%%%%%%%%%%%%%%%%%%%%%%%%%%%%%

%\begin{figure}
%\centering
%\includegraphics[width=7.5cm]{comparison_max_rate_min_rate}
%\caption{Comparison of achievable rates of transmitters between the SRM and MRM policies.}
%\label{fig:comparison_max_rate_min_rate}
%\end{figure}
%\begin{figure}
%\centering
%\includegraphics[width=7.5cm]{R_1_R_2_ratio_vs_slots}
%\caption{Ratio of average achievable rates, $R_1/R_2$ versus the total number of slots ($M$) for the MRM policy with perfect CSI.}
%\label{fig:R_1_R_2_ratio_vs_slots}
%\end{figure}
For the SRM and MRM policies with perfect CSI, Fig. \ref{fig:comparison_max_rate_min_rate} shows the average achievable sum-rate, the rate of Tx$_1$ ($R_1$), and the rate of Tx$_2$ ($R_2$) as a function of total number of slots ($M$) averaged over 300 channel realizations. For both the policies, the average achievable rate increases with the number of slots as expected. However, the unfair achievable rates in SRM policy can be observed from the figure. The farther transmitter Tx$_2$ achieves a much smaller rate than that by the transmitter Tx$_1$ due to higher path loss. The MRM policy, on the other hand, improves the user fairness at the cost of reduced sum-rate as shown in Fig. \ref{fig:R_1_R_2_ratio_vs_slots}.

From Fig. \ref{fig:comparison_max_rate_min_rate}, observe that the rate achieved by Tx$_1$ is reduced and the rate achieved by Tx$_2$ is increased significantly in the MRM policy. This is because the MRM policy allows Tx$_2$ to harvest for more number of slots and transmit with higher power while restricting the transmit power of Tx$_1$ at the same time.

\textit{\textbf{Remark:}} Ideally, the optimization problem in (\ref{eq:min_rate_convex_obj})-(\ref{eq:min_rate_convex_c2}) should result in $R_1=R_2=\bar{R}$, but this is not the case (as shown in Fig. \ref{fig:R_1_R_2_ratio_vs_slots}) if $M$ is small. This is because an entire slot is allocated to either of the transmitter for the transmission, and therefore $R_1$ and $R_2$ may not always be equal. However, as we increase $M$, the achievable rates by both transmitters become equal asymptotically as shown in Fig. \ref{fig:R_1_R_2_ratio_vs_slots}.
\subsubsection{Effect of path loss exponent ($\alpha$)}
\begin{figure*}[h]
	\centering
	\begin{minipage}{0.48\linewidth}
		\centering
\includegraphics[width=0.9\linewidth]{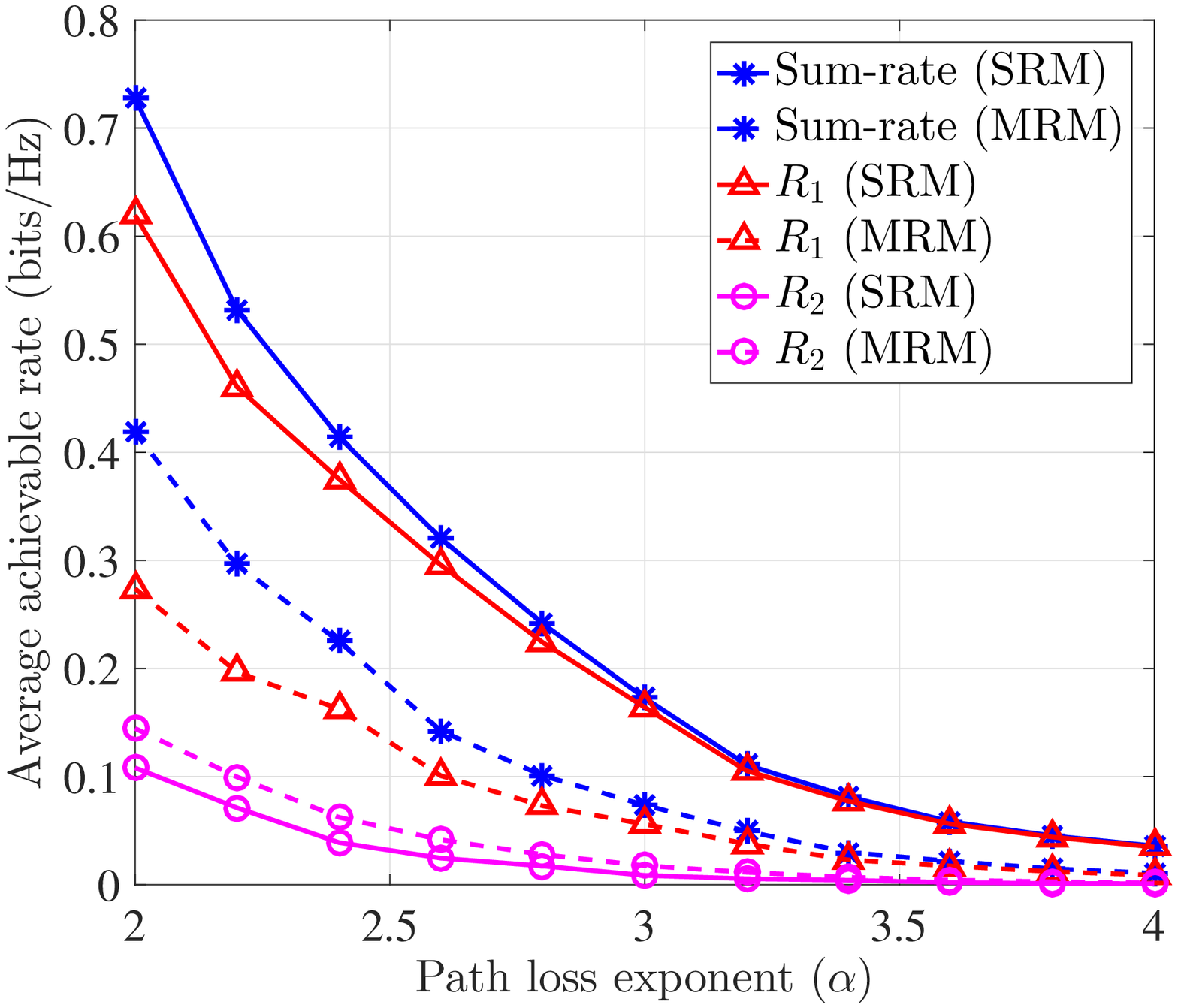}
\caption{Comparison of the effect of the path loss exponent $\alpha$ on rates achieved by both transmitters, $M=4$.}
\label{fig:throughput_vs_path_loss_comparison}
	\end{minipage}
	\hspace{3mm}
	\begin{minipage}{0.48\linewidth}
		\centering
\includegraphics[width=0.9\linewidth]{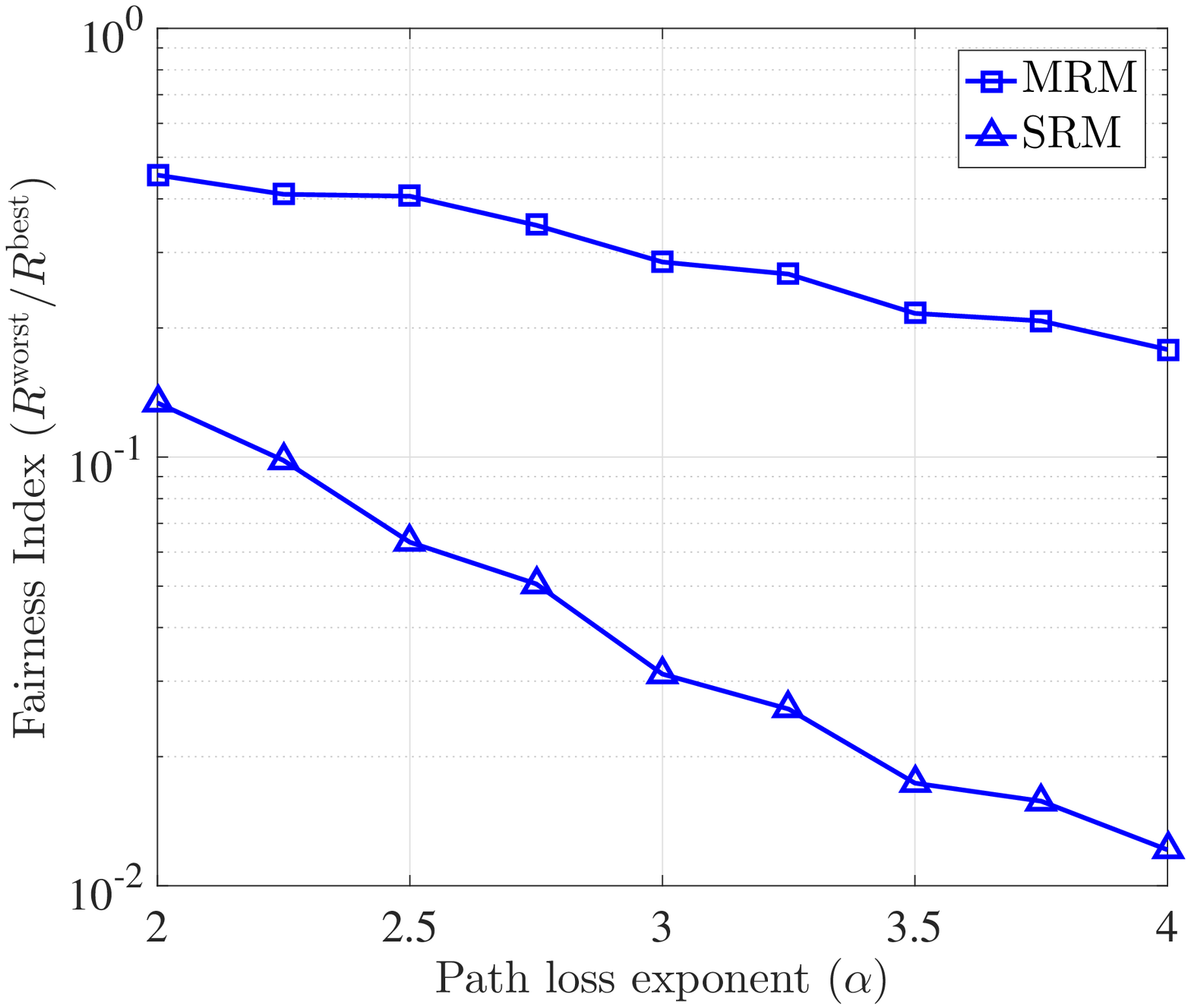}
\caption{Fairness index $\mathcal{F}$ versus the path loss exponent $\alpha$ for SRM and MRM policies with perfect CSI, $M=4$.}
\label{fig:R_2_R_1_ratio_comparison}
	\end{minipage}
\end{figure*}
\begin{figure*}[h]
	\centering
	\begin{minipage}{0.48\linewidth}
		\centering
		\includegraphics[width=0.9\linewidth]{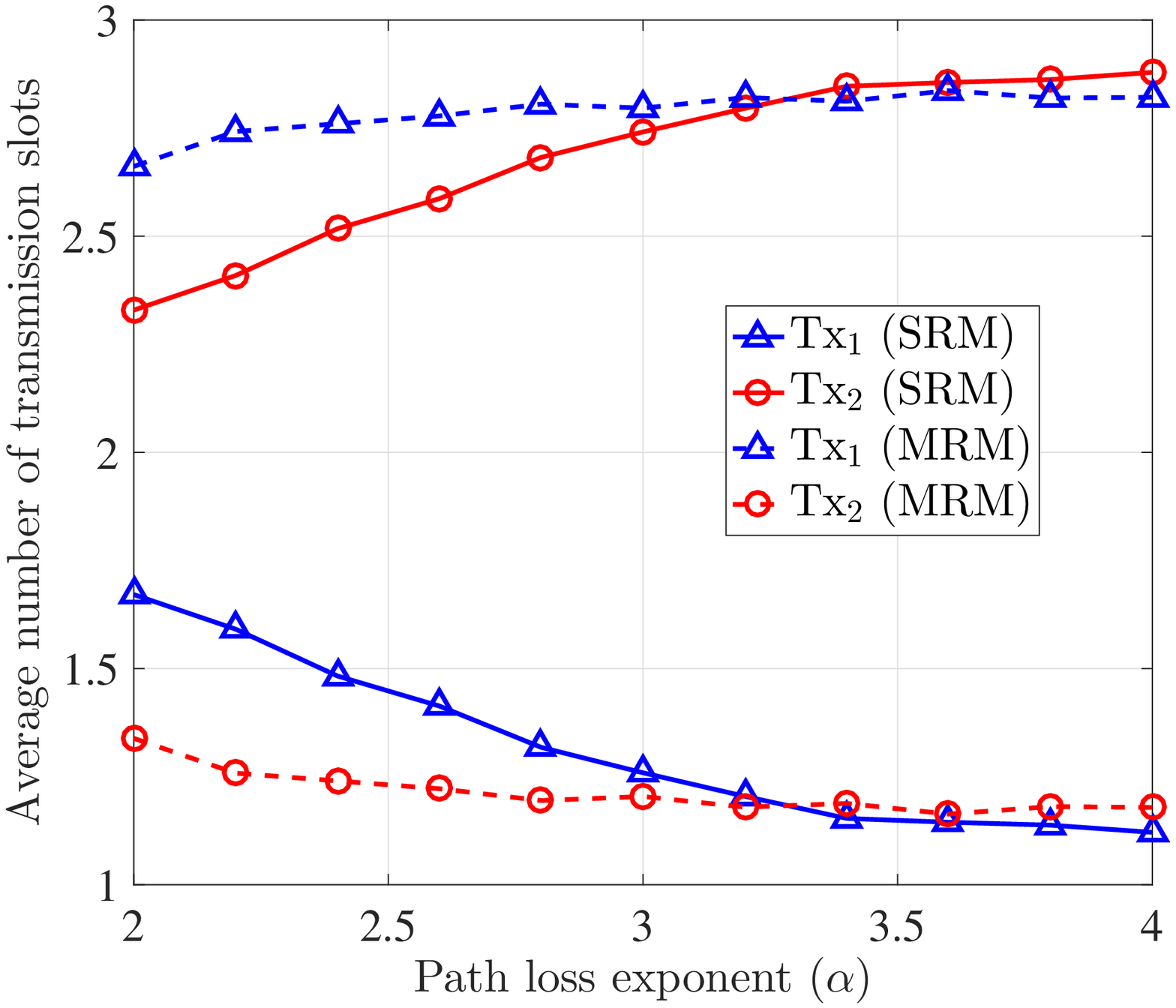}
		\caption{Comparison of the effect of the path loss exponent $\alpha$ on average number of transmission slots, $M=4$.}
		\label{fig:slot_allocation_vs_path_loss_comparison}
	\end{minipage}
	\hspace{3mm}
	\begin{minipage}{0.48\linewidth}
		\centering
		\includegraphics[width=0.9\linewidth]{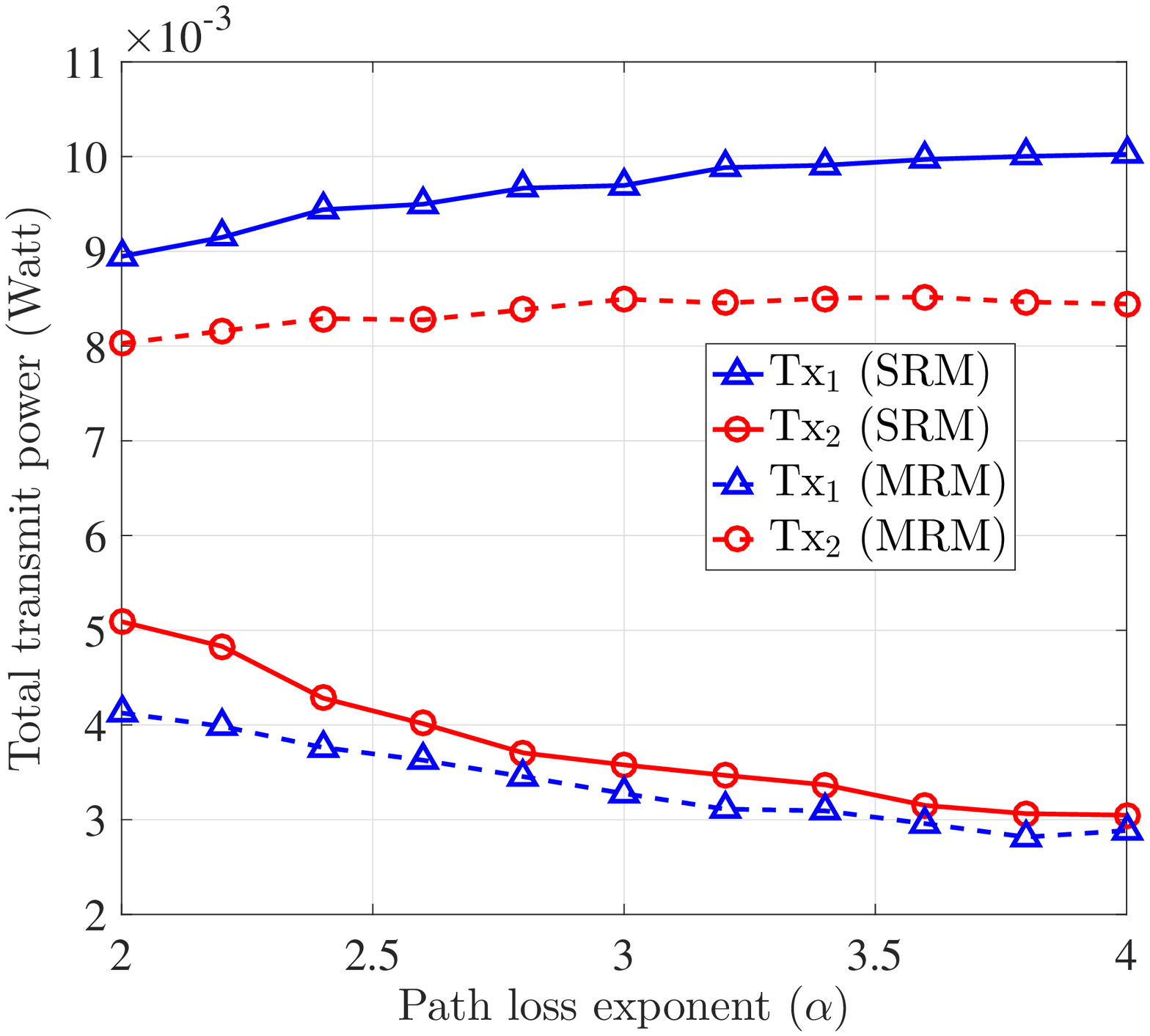}
		\caption{Comparison of the effect of the path loss exponent $\alpha$ on transmit power, $M=4$.}
		\label{fig:total_power_vs_path_loss_comparison}
	\end{minipage}
\end{figure*}
Fig. \ref{fig:throughput_vs_path_loss_comparison} shows the average achievable rates as a function of $\alpha$ for both the policies. Observe that as $\alpha$ increases, the average achievable rate reduces exponentially as $h_{k,i}\propto D_i^{-\alpha}$. However the effect of increasing $\alpha$ is more severe on the farther transmitter Tx$_2$ as $D_2 > D_1$, and thus $h_{2,i}$ (includes the path loss) reduces more rapidly than $h_{1,i}$. Also, as $\alpha$ increases, the rate achieved by Tx$_2$ approaches to zero.

The degree of unfairness caused by the SRM policy can be observed by Fig. \ref{fig:R_2_R_1_ratio_comparison}. We can measure the fairness by defining a fairness index as $\mathcal{F} = \frac{R^{\rm{worst}}}{R^{\rm{best}}}$, where $R^{\rm{worst}}$ and $R^{\rm{best}}$ represent the rates achieved by the worst and the best transmitter, respectively. A higher value of $\mathcal{F}$ represents more fairness. In case of $K=2$, we have $R^{\rm{best}}=R_1$ and $R^{\rm{worst}}=R_2$. It is observed that as the path loss exponent increases, the rate achieved by the farther transmitter reduces significantly. When $\alpha=4$, we have $\C{F}\approx0.012$, which means that the transmitter closer to the receiver achieves a rate that is 82$\times$ the rate achieved by the farther transmitter. On the other hand, the MRM policy introduces more fairness in the system, which can be observed from Figs. \ref{fig:R_1_R_2_ratio_vs_slots} and \ref{fig:R_2_R_1_ratio_comparison} as $\mathcal{F}_{\rm{MRM}}>\mathcal{F}_{\rm{SRM}}$. Specifically, when $\alpha=4$, we have $\C{F}\approx0.18$, which means that the transmitter closer to the receiver achieves a rate that is 5.5$\times$ the rate achieved by the farther transmitter. In addition, Fig. \ref{fig:R_2_R_1_ratio_comparison} shows that the fairness reduces with $\alpha$, which is more severe for SRM policy as the slope of fairness index $\mathcal{F}$ in the SRM policy is higher than the slope in the MRM policy.
%\begin{figure*}[h]
%	\centering
%	\begin{minipage}{0.48\linewidth}
%		\centering
%\includegraphics[width=0.9\linewidth]{slot_allocation_vs_path_loss_comparison}
%\caption{Comparison of the effect of the path loss exponent $\alpha$ on average number of transmission slots, $M=4$.}
%\label{fig:slot_allocation_vs_path_loss_comparison}
%	\end{minipage}
%	\hspace{3mm}
%	\begin{minipage}{0.48\linewidth}
%		\centering
%\includegraphics[width=0.9\linewidth]{total_power_vs_path_loss_comparison}
%\caption{Comparison of the effect of the path loss exponent $\alpha$ on transmit power, $M=4$.}
%\label{fig:total_power_vs_path_loss_comparison}
%	\end{minipage}
%\end{figure*}

Figs. \ref{fig:slot_allocation_vs_path_loss_comparison} and \ref{fig:total_power_vs_path_loss_comparison} show the effect of path loss exponent on slot allocation and transmit powers, respectively, and are obtained for 500 channel realizations with $M=4$. In Fig. \ref{fig:slot_allocation_vs_path_loss_comparison}, observe that for a fixed $\alpha$, the SRM policy allocates less number of transmission slots to Tx$_1$ than Tx$_2$, whereas the MRM policy allocates more number of transmission slots to Tx$_1$ than Tx$_2$. In addition, Fig. \ref{fig:total_power_vs_path_loss_comparison} shows that the total transmit power of Tx$_1$ is higher than that of the Tx$_2$ in SRM policy for fixed $\alpha$ whereas, the trend is opposite in MRM policy. This joint slot allocation and transmit power control of MRM policy introduces fairness among users. Since the number of transmission slots for Tx$_2$ are less in MRM policy, Tx$_2$ now harvests more energy and transmits with higher power. This results in higher achievable rate $R_2$ than that for the SRM policy. On the other hand, the number of transmission slots for Tx$_1$ are increased in MRM policy, hence Tx$_1$ harvests less energy and transmits with a lower power. This results in a smaller achievable rate than that for the SRM policy.

In Fig. \ref{fig:slot_allocation_vs_path_loss_comparison}, observe that as $\alpha$ increases, in the SRM (MRM) policy, the number of transmission slots allocated to Tx$_1$ decreases (increases) and the number of transmission slots allocated to Tx$_2$ increases (decreases). Also Fig. \ref{fig:total_power_vs_path_loss_comparison} shows that the transmit power of Tx$_1$ increases (decreases) and the transmit power of Tx$_2$ decreases (increases) in the SRM (MRM) policy. This results in reduced achievable rates for both transmitters in both the policies.
\subsubsection{Effect of number of transmitters ($K$)}
\begin{figure}[!h]
\centering
\includegraphics[width=0.9\linewidth]{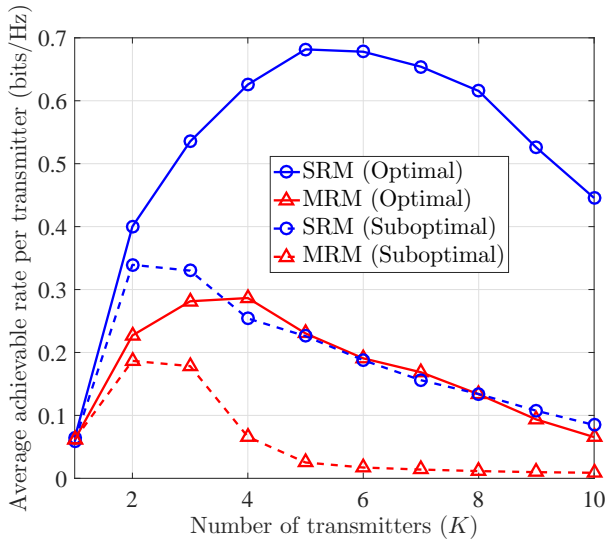}
\caption{Average achievable rate per transmitter versus the number of  transmitter ($K$) with the perfect CSI, $M=4$.}
\label{fig:throughput_vs_users_max_rate_full_CSI}
\end{figure}
\begin{figure}[!h]
			\centering
	\includegraphics[width=0.9\linewidth]{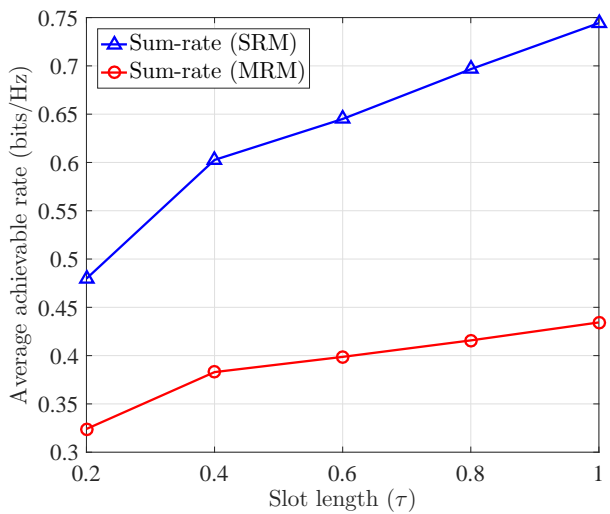}
	\caption{Average achievable sum-rates versus slot length ($\tau$) under the SRM and MRM policies.}
	\label{fig:throughput_vs_slot_length}
\end{figure}
\begin{figure}[!h]
		\centering
		\includegraphics[width=0.9\linewidth]{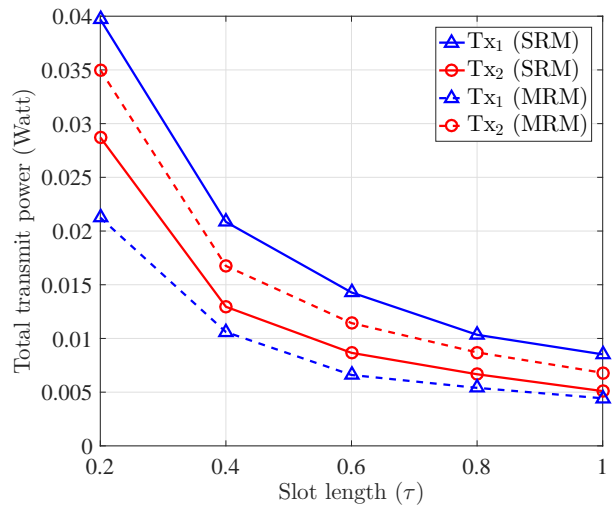}
		\caption{Average total transmit power versus slot length ($\tau$) under the SRM and MRM policies.}
		\label{fig:transmit_pwr_vs_slot_length}
\end{figure}
\begin{figure}[!h]
			\centering
	\includegraphics[width=0.9\linewidth]{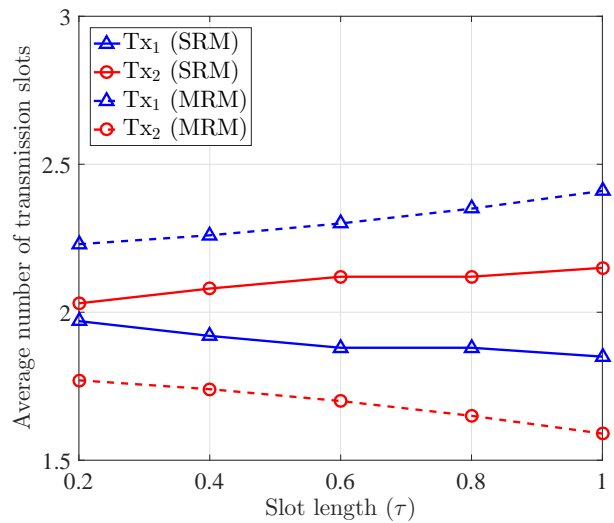}
	\caption{Average number of transmission slots versus slot length ($\tau$) under the SRM and MRM policies.}
	\label{fig:tx_slots_vs_slot_length}
\end{figure}
Fig. \ref{fig:throughput_vs_users_max_rate_full_CSI} shows the effect of number of transmitters on the average achievable rate per transmitter for $M=4$ slots for the SRM and MRM policies. It is assumed that the $k$th transmitter is located at distance $D_k$ from the receiver such that $D_k=\frac{D_K}{K}\times k, \; \forall k$, where $D_K=10$ m. The energy available for all transmitters in all the slots is assumed to be the same ($E_H^{k,i}=3$ mJ, $\forall k,i$). The path loss exponent $\alpha$ is fixed to 2. For $K=1$, the performance of both policies is the same. For the SRM policy, observe that the achievable rate per transmitter initially increases with the number of transmitters and then decreases. This is because initially all slots may not be utilized efficiently. As the number of transmitters increases, the achievable rate per transmitter increases as the number of slots are sufficient to schedule them optimally, and even the weakest transmitter gets a chance to transmit. However, if we further increase the number of transmitters, the network becomes overcrowded, and weaker transmitters may not get scheduled at all (as they have poor channel due to large path loss). Hence the achievable rate per transmitter reduces.

An increase in the number of transmitters has a negative effect for the MRM policy. This is because as the number of transmitter increases, the farther transmitters also get scheduled as the scheduler aims to maximize the minimum-rate in the network. Therefore, some of  the harvesting/transmission slots allocated to transmitters closer to the receiver are reduced and assigned to farther transmitters in order to satisfy their target rate. Due to this reassignment, the rates of closer transmitters reduce, and since the farther transmitters have poor channel conditions due to large path loss, their maximum achievable rate reduces. Hence the achievable rate per transmitter reduces significantly as we increase the number of transmitters.

\subsubsection{Effect of slot length ($\tau$)}
 Fig. \ref{fig:throughput_vs_slot_length} shows the effects of slot length $\tau$ on the average achievable rate for both the policies. In Fig. \ref{fig:throughput_vs_slot_length}, observe that, as the $\tau$ increases, the average achievable sum-rate in both policies increases due to the relation $R_{\rm sum}=\sum_{i=1}^M\sum_{k=1}^K\tau\log_2\left(1+\frac{|g_{k,i}|^2 E_C^{k,i}}{\tau\sigma_n^2}\right)$, where $E_C^{k,i}$ is the energy consumed by the $k$th transmitter in $i$th slot. Also as Fig. \ref{fig:transmit_pwr_vs_slot_length} shows, the total transmit power of the transmitters decreases with slot length. This is because the energy availability at all transmitters is the same for each $\tau$, and as $\tau$ increases the transmit power decreases due to the relation $P=\frac{E}{\tau}$. Fig. \ref{fig:tx_slots_vs_slot_length} shows that the average number of transmission slots for Tx$_1$ (Tx$_2$) decreases (increases) for the SRM problem with the increase in $\tau$. Similarly the number of transmission slots for Tx$_1$ (Tx$_2$) increases (decreases) for the MRM problem with the increase in $\tau$.

\subsection{Suboptimal Policy with Perfect CSI}
\begin{figure}[!h]
\centering
\includegraphics[width=7.5cm]{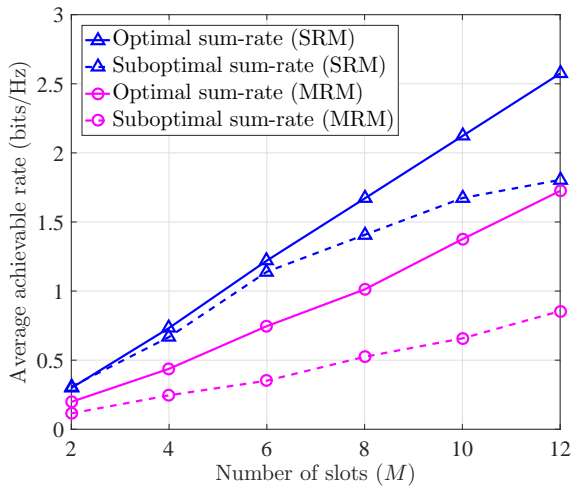}
\caption{Comparison of optimal and low-complexity suboptimal SRM and MRM policies.}
\label{fig:throughput_vs_slots_full_CSI_comparison_heuristic}
\end{figure}
Fig. \ref{fig:throughput_vs_slots_full_CSI_comparison_heuristic} shows the performance of the optimal and the low-complexity suboptimal algorithms for two users under the assumption of perfect CSI. Although the suboptimal algorithm does not perform well as compared to the optimal policy, its complexity is much less as it only solves a convex optimization problem twice.

Fig. \ref{fig:throughput_vs_users_max_rate_full_CSI} shows the effect of number of transmitters on the average achievable rate per transmitter for the suboptimal policies. As we increase the number of transmitters ($K$), the average achievable rate per transmitter first increases and then reduces rapidly as in optimal policies. The reason of this rapid decay is that, as we increase $K$ beyond 2, the optimal solution of the relaxed problem results in $w^{\rm rel}_{k,i}<0.5$ for most $(k,i)$. In this case, rounding off these $w^{\rm rel}_{k,i}$ results in $w^{\rm round}_{k,i}=0$, which makes the transmitters harvest energy in most of the slots and reduces the average achievable rate per transmitter. Also, in some slots it may happen that none of the transmitters transmit and all of them harvest energy.
\subsection{Optimal Policy with Imperfect CSI}
%\begin{figure*}[!ht]
%	\begin{minipage}{0.48\linewidth}
%		\centering
%\includegraphics[width=0.9\linewidth]{throughput_vs_M_and_path_loss_diff_epsilon_comparison1}
%\caption{Average achievable sum-rate versus the number of slots $M$ for SRM and MRM policies under the bounded channel estimation error.}
%\label{fig:max_rate_throughput_vs_M_and_slots_diff_epsilon_comparison1}
%	\end{minipage}\hspace{4mm}
%	\begin{minipage}{0.48\linewidth}
%		\centering
%	\includegraphics[width=0.9\linewidth]{comparison_with_full_duplex_and_zhang}
%\caption{Comparison of sum-rates under the SRM and MRM policies.}
%\label{fig:comparison_with_literature_SR}
%	\end{minipage}
%\end{figure*}
%
%\begin{figure}
%\centering
%\includegraphics[width=7.5cm]{throughput_vs_M_and_path_loss_diff_epsilon_comparison1}
% \caption{Average achievable sum-rate versus the number of slots $M$ for both the policies under the bounded channel estimation error.}
%\label{fig:max_rate_throughput_vs_M_and_slots_diff_epsilon_comparison1}
%\end{figure}
In this subsection we present simulation results for both the optimal policies with imperfect CSI at the receiver. We obtained the results for the worst-case scenario. We assume the estimated channel gain $\hat{g}_{k,i}\sim \C{N}(0,1)$.
\subsubsection{Effect of Uncertainty Bound $\epsilon$}
\begin{figure}[!h]
		\centering
\includegraphics[width=0.9\linewidth]{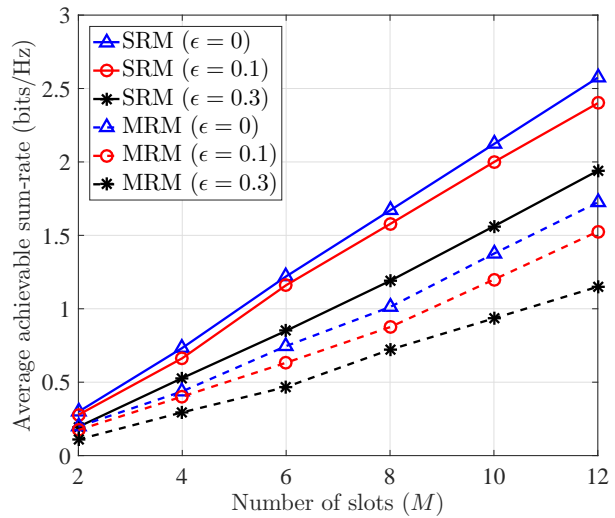}
\caption{Average achievable sum-rate versus the number of slots $M$ for SRM and MRM policies under the bounded channel estimation error.}
\label{fig:max_rate_throughput_vs_M_and_slots_diff_epsilon_comparison1}
\end{figure}
Fig. \ref{fig:max_rate_throughput_vs_M_and_slots_diff_epsilon_comparison1} shows the average achievable sum-rate versus the number of slots $M$ for different values of channel uncertainty bound $\epsilon$ under both the policies. The effects of all other parameters except $\epsilon$ is the same as in the case of perfect CSI. As $\epsilon$ increases, the worst-case rate reduces, and hence the average achievable sum rate under the SRM policy. However, when we solve the problem (\ref{eq:min_rate_iCSI_obj})-(\ref{eq:min_rate_iCSI_c2}), the feasible set becomes more stringent than that in (\ref{eq:min_rate_orig_obj})-(\ref{eq:min_rate_orig_c2}), which results in the reduced sum-rate for the MRM policy.
\subsection{Comparison with Myopic Policies of \cite{TDMA_zhang,TDMA_full_duplex_kang}}
\label{subsec:comparison_myopic_result}
\begin{figure}[!h]
		\centering
\includegraphics[width=0.9\linewidth]{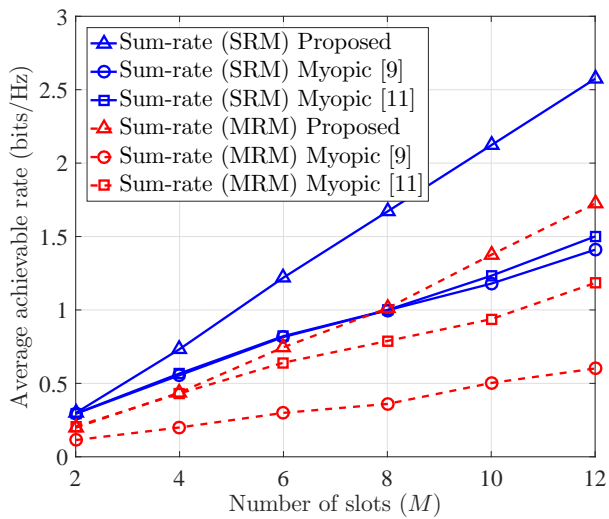}
\caption{Comparison of sum-rates under the SRM and MRM policies.}
\label{fig:comparison_with_literature_SR}
\end{figure}
\begin{figure}[!h]
	\centering
	\includegraphics[width=0.95\linewidth]{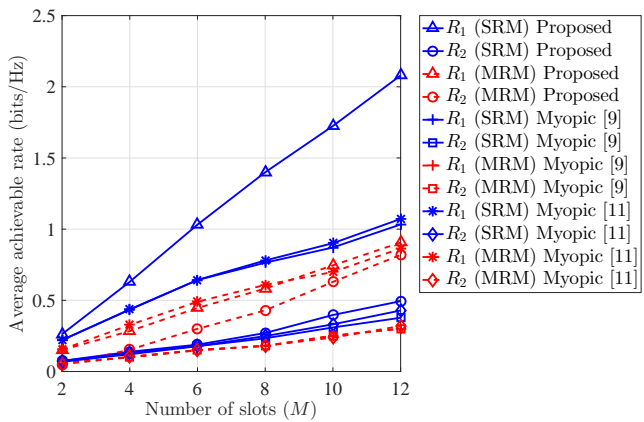}
	\caption{Comparison of rates achieved by Tx$_1$ and Tx$_2$ under SRM and MRM policies.}
	\label{fig:comparison_with_literature_R1}
\end{figure}
Fig. \ref{fig:comparison_with_literature_SR} shows the comparison of achievable sum-rate under the SRM and MRM problems for different policies. In Fig. \ref{fig:comparison_with_literature_SR}, observe that both the proposed SRM and MRM policies achieve higher sum-rate than that for myopic policies proposed in  \cite{TDMA_zhang} and \cite{TDMA_full_duplex_kang}. This is because our proposed policies optimize the scheduling over all time slots jointly, rather than optimizing over each slot separately as in  \cite{TDMA_zhang} and \cite{TDMA_full_duplex_kang}. This allows a better utilization of the energy given the channel gains.
	
	In addition, in Fig. \ref{fig:comparison_with_literature_R1}, observe that the proposed MRM policy offers a significant amount of fairness among the transmitters as compared to the MRM policy based on \cite{TDMA_full_duplex_kang}. Although the MRM policy based on \cite{TDMA_zhang} gives the best fairness by ensuring $R_1=R_2$, it reduces the achievable sum-rate of the network, which can be observed in Fig. \ref{fig:comparison_with_literature_SR}. Our proposed MRM policy gives a better sum-rate than the MRM policies of \cite{TDMA_zhang} and \cite{TDMA_full_duplex_kang}. As discussed for Fig. \ref{fig:R_1_R_2_ratio_vs_slots} that our proposed MRM policy achieves the best fairness, \textit{i.e.}, $R_1=R_2$ for sufficiently high number of slots. Thus our MRM policy performs better than the one in \cite{TDMA_zhang} in terms of fairness and sum-rate for higher values of $M$.

\textbf{Complexity of the proposed policy:} The GBD algorithm solves a convex optimization and an MILP problem in each iteration. The MILPs can be solved using Branch and Bound algorithms, which are NP-hard and have exponential complexity \cite{BnB_complexity}. The convex optimization problem on the other hand, can be solved in polynomial time. The myopic policies in \cite{TDMA_zhang} and \cite{TDMA_full_duplex_kang} are convex optimization problems and can be solved in polynomial time. In our proposed policies, we are gaining in terms of achievable throughput at the cost of increased complexity. 

\section{Future Directions}
\label{sec:future}
%In this work, we have assumed the battery capacity of all the transmitters to be infinite. This assumption can be justified in scenarios where the amount of energy harvested from the environment is very small compared to the battery capacity. In such cases, the battery is large enough to store all the harvested energy packets and battery overflow does not occur. In this case, the system performance is same as if we had an infinite capacity battery \cite{infinite_battery_velkov}. Also, assuming infinite capacity battery gives an \textit{upper bound} on the system performance for any practical implementation \cite{energy_allocation_zhang}.
We now discuss a few future directions of our work:
	\begin{enumerate}
		\item \textbf{Effects of finite battery:} In our system model, the energy arrival  is random, and in each slot, a transmitter decides whether to store arriving energy in its battery or transmit its data. If transmitters have finite capacity batteries and the arrived energy is more than what they can store, the excess energy would not be saved in the battery and would get wasted.  Moreover, if a transmitter is sending its data, the arrived energy is not stored in the battery and gets wasted. Hence the effects of finite capacity battery on the achievable rates is worth investigating.
		\item \textbf{Energy wastage minimization:} The finite battery capacity results in wastage of harvested energy due to battery overflow. Thus it is interesting to investigate a new problem where we wish to maximize the throughput while keeping the energy wastage below an acceptable threshold.
	\end{enumerate}
\section{Conclusions}
\label{sec:conclusions}
We considered an energy harvesting network where multiple EH transmitters have random energy arrivals and communicate with the common receiver in a time sharing basis. We assume a slotted mode of operation. The transmitters employ a \textit{harvest-or-transmit} protocol, \textit{i.e.}, in each slot, a transmitter can either harvest energy from the environment or transmits its data to the receiver. Under these settings, we obtained an optimal slot allocation and power control policy maximizing the sum-rate of all transmitters using the GBD algorithm. We observed that this policy results in an unfair rate allocation among transmitters. To induce fairness, we considered a problem of maximizing the minimum rate in the network. This policy improves the fairness by assigning energy harvesting slots of strong transmitters to weak transmitters and thereby increasing their transmit power. However, this results in the reduced sum-rate due the large path loss for farther transmitters. We observed that both transmission policies are greatly affected by the path loss exponent. We also compared the proposed policies with myopic policies proposed in the literature and showed that the proposed policies outperform myopic policies in terms of achievable rates.

We also considered the case of the imperfect CSI at the receiver and obtained the robust SRM and MRM policies. We investigated the effects of the radius of uncertainty region $\epsilon$ on the optimal policies, and it is observed that as $\epsilon$ increases, the achievable rates decrease. We also proposed a low-complexity suboptimal algorithm for both the SRM and MRM problems. Although both the suboptimal policies underperform, their computational complexities are much smaller than that of the optimal policies.

%\appendices
%% you can choose not to have a title for an appendix
%% if you want by leaving the argument blank
%\section{}
%% use section* for acknowledgment
%\section*{Acknowledgment}


\ifCLASSOPTIONcaptionsoff
  \newpage
\fi

\bibliographystyle{ieeetr} %abbrv, acm, alpha
\bibliography{references}

\begin{thebibliography}{10}

\bibitem{EH_survey_Ku}
M.~L. Ku, W.~Li, Y.~Chen, and K.~J.~R. Liu, ``Advances in energy harvesting
  communications: {P}ast, present, and future challenges,'' {\em IEEE
  Communications Surveys \& Tutorials}, vol.~18, pp.~1384--1412, Second quarter
  2016.

\bibitem{EH_survey_1}
Y.~He, X.~Cheng, W.~Peng, and G.~L. Stuber, ``A survey of energy harvesting
  communications: Models and offline optimal policies,'' {\em IEEE
  Communications Magazine}, vol.~53, pp.~79--85, June 2015.

\bibitem{EH_survey_2}
X.~Lu, P.~Wang, D.~Niyato, D.~I. Kim, and Z.~Han, ``Wireless networks with {RF}
  energy harvesting: A contemporary survey,'' {\em IEEE Communications Surveys
  \& Tutorials}, vol.~17, pp.~757--789, Second quarter 2015.

\bibitem{EH_survey_Sudevalayam}
S.~Sudevalayam and P.~Kulkarni, ``Energy harvesting sensor nodes: Survey and
  implications,'' {\em IEEE Communications Surveys \& Tutorials}, vol.~13,
  pp.~443--461, Third quarter 2011.

\bibitem{swipt_magzine_krikidis}
I.~Krikidis, S.~Timotheou, S.~Nikolaou, G.~Zheng, D.~W.~K. Ng, and R.~Schober,
  ``Simultaneous wireless information and power transfer in modern
  communication systems,'' {\em IEEE Communications Magazine}, vol.~52,
  pp.~104--110, November 2014.

\bibitem{TCTM_ulukus}
J.~Yang and S.~Ulukus, ``Transmission completion time minimization in an energy
  harvesting system,'' in {\em 44th Annual Conference on Information Sciences
  and Systems (CISS)}, (Princeton, NJ), pp.~1--6, March 2010.

\bibitem{STTM_yener}
K.~Tutuncuoglu and A.~Yener, ``Short-term throughput maximization for battery
  limited energy harvesting nodes,'' in {\em IEEE International Conference on
  Communications (ICC)}, (Kyoto, Japan), pp.~1--5, June 2011.

\bibitem{STTM_TCTM_fading}
O.~Ozel, K.~Tutuncuoglu, J.~Yang, S.~Ulukus, and A.~Yener, ``Transmission with
  energy harvesting nodes in fading wireless channels: Optimal policies,'' {\em
  IEEE Journal on Selected Areas in Communications}, vol.~29, pp.~1732--1743,
  September 2011.

\bibitem{TDMA_zhang}
H.~Ju and R.~Zhang, ``Throughput maximization in wireless powered communication
  networks,'' {\em IEEE Transactions on Wireless Communications}, vol.~13,
  pp.~418--428, January 2014.

\bibitem{TDMA_velkov}
Z.~Hadzi-Velkov, I.~Nikoloska, G.~K. Karagiannidis, and T.~Q. Duong, ``Wireless
  networks with energy harvesting and power transfer: Joint power and time
  allocation,'' {\em IEEE Signal Processing Letters}, vol.~23, pp.~50--54,
  January 2016.

\bibitem{TDMA_full_duplex_kang}
X.~Kang, C.~K. Ho, and S.~Sun, ``Full-duplex wireless-powered communication
  network with energy causality,'' {\em IEEE Transactions on Wireless
  Communications}, vol.~14, pp.~5539--5551, October 2015.

\bibitem{TDMA_underlay_CRN}
D.~Xu and Q.~Li, ``Joint power control and time allocation for wireless powered
  underlay cognitive radio networks,'' {\em IEEE Wireless Communications
  Letters}, vol.~6, pp.~294--297, June 2017.

\bibitem{full_duplex_zhang}
H.~Ju and R.~Zhang, ``Optimal resource allocation in full-duplex
  wireless-powered communication network,'' {\em IEEE Transactions on
  Communications}, vol.~62, pp.~3528--3540, October 2014.

\bibitem{TDMA_vs_NOMA}
Q.~Wu, W.~Chen, D.~W.~K. Ng, and R.~Schober, ``Spectral and energy efficient
  wireless powered {I}o{T} networks: {NOMA} or {TDMA}?,'' {\em IEEE
  Transactions on Vehicular Technology}, accepted.

\bibitem{imperfect_CSI_relay_ahmed}
I.~Ahmed, A.~Ikhlef, D.~W.~K. Ng, and R.~Schober, ``Optimal resource allocation
  for energy harvesting two-way relay systems with channel uncertainty,'' in
  {\em IEEE Global Conference on Signal and Information Processing
  (GlobalSIP)}, (Austin, TX), pp.~345--348, December 2013.

\bibitem{imperfect_CSI_relay_half_duplex_ahmed}
I.~Ahmed, A.~Ikhlef, D.~W.~K. Ng, and R.~Schober, ``Power allocation for a
  hybrid energy harvesting relay system with imperfect channel and energy state
  information,'' in {\em IEEE Wireless Communications and Networking Conference
  (WCNC)}, (Istanbul, Turkey), pp.~990--995, April 2014.

\bibitem{imperfect_CSI_CRN}
S.~Gong, L.~Duan, and P.~Wang, ``Robust optimization of cognitive radio
  networks powered by energy harvesting,'' in {\em IEEE Conference on Computer
  Communications (INFOCOM)}, (Hong Kong), pp.~612--620, April 2015.

\bibitem{robust_RA_MIMO}
E.~Boshkovska, D.~W.~K. Ng, N.~Zlatanov, A.~Koelpin, and R.~Schober, ``Robust
  resource allocation for {MIMO} wireless powered communication networks based
  on a non-linear {EH} model,'' {\em IEEE Transactions on Communications},
  vol.~65, pp.~1984--1999, May 2017.

\bibitem{robust_transceiver_MIMO_qin}
J.~Xiao, C.~Xu, Q.~Zhang, X.~Huang, and J.~Qin, ``Robust transceiver design for
  two-user {MIMO} interference channel with simultaneous wireless information
  and power transfer,'' {\em IEEE Transactions on Vehicular Technology},
  vol.~65, pp.~3823--3828, May 2016.

\bibitem{robust_transceiver_swipt_wang}
T.~Peng, F.~Wang, Y.~Huang, and X.~Wang, ``Robust transceiver optimization for
  {MISO SWIPT} interference channel: {A} decentralized approach,'' in {\em IEEE
  Vehicular Technology Conference (VTC Spring)}, (Nanjing, China), pp.~1--5,
  May 2016.

\bibitem{save_then_transmit_zhang}
S.~Luo, R.~Zhang, and T.~J. Lim, ``Optimal save-then-transmit protocol for
  energy harvesting wireless transmitters,'' {\em IEEE Transactions on Wireless
  Communications}, vol.~12, pp.~1196--1207, March 2013.

\bibitem{convex_MINLP}
P.~Bonami, M.~Kilin{\c{c}}, and J.~Linderoth, {\em Algorithms and Software for
  Convex Mixed Integer Nonlinear Programs}, pp.~1--39.
\newblock New York, NY: Springer New York, 2012.

\bibitem{GBD}
A.~M. Geoffrion, ``Generalized {B}enders decomposition,'' {\em Journal of
  Optimization Theory and Applications}, vol.~10, no.~4, pp.~237--260, 1972.

\bibitem{channel_estimation_survey_2}
H.~Arslan and G.~E. Bottomley, ``Channel estimation in narrowband wireless
  communication systems,'' {\em Wireless Communications and Mobile Computing},
  vol.~1, no.~2, pp.~201--219, 2001.

\bibitem{fading_prediction}
A.~Duel-Hallen, ``Fading channel prediction for mobile radio adaptive
  transmission systems,'' {\em Proceedings of the IEEE}, vol.~95,
  pp.~2299--2313, December 2007.

\bibitem{pilot_based_prediction_1}
T.~Ekman, {\em Prediction of Mobile Radio Channels: Modeling and Design}.
\newblock PhD thesis, Uppsala University, Uppsala, Sweden, 2002.

\bibitem{channel_estimation_survey}
J.~Wu and P.~Fan, ``A survey on high mobility wireless communications:
  Challenges, opportunities and solutions,'' {\em IEEE Access}, vol.~4,
  pp.~450--476, 2016.

\bibitem{info_theory_book}
T.~M. Cover and J.~A. Thomas, {\em Elements of Information Theory (Wiley Series
  in Telecommunications and Signal Processing)}.
\newblock Wiley-Interscience, 2006.

\bibitem{cvx_book}
S.~Boyd and L.~Vandenberghe, {\em Convex Optimization}.
\newblock New York, NY, USA: Cambridge University Press, 2004.

\bibitem{MINLP_floudas}
C.~A. Floudas, {\em Nonlinear and Mixed-Integer Optimization}.
\newblock New York, NY, USA: Oxford University Press, 1995.

\bibitem{Benders}
J.~F. Benders, ``Partitioning procedures for solving mixed-variables
  programming problems,'' {\em Numerische Mathematik}, vol.~4, 1962.

\bibitem{cvx}
M.~Grant and S.~Boyd, ``{CVX}: Matlab software for disciplined convex
  programming, version 2.1.'' http://cvxr.com/cvx, March 2014.

\bibitem{dual_descent}
A.~Nedi\'{c} and A.~Ozdaglar, ``Approximate primal solutions and rate analysis
  for dual subgradient methods,'' {\em SIAM Journal on Optimization}, vol.~19,
  no.~4, pp.~1757--1780, 2009.

\bibitem{bedi_tcom}
A.~S. Bedi and K.~Rajawat, ``Network resource allocation via stochastic
  subgradient descent: Convergence rate,'' {\em IEEE Transactions on
  Communications}, vol.~PP, no.~99, pp.~1--1, 2018.

\bibitem{mosek}
``{MOSEK} {S}oftware.'' http://www.mosek.com.

\bibitem{BnB_complexity}
J.~Till, S.~Engell, S.~Panek, and O.~Stursberg, ``Empirical complexity analysis
  of a {MILP}-approach for optimization of hybrid systems,'' {\em IFAC
  Proceedings Volumes}, vol.~36, pp.~129 -- 134, June 2003.

\end{thebibliography}

\end{document}